\documentclass[11pt]{article}

\usepackage[T1]{fontenc}
\usepackage{amsmath}
\usepackage{newtxtext,newtxmath}
\usepackage[letterpaper,top=0.88in,bottom=0.90in,
            left=1in,right=1in,headheight=22pt]{geometry}
\usepackage{booktabs}
\usepackage{graphicx}
\usepackage[table]{xcolor}
\usepackage{colortbl}
\usepackage{array}
\usepackage{makecell}
\usepackage{enumitem}
\usepackage{microtype}
\usepackage{float}
\usepackage{setspace}
\usepackage{abstract}
\usepackage{titlesec}
\usepackage{needspace}
\usepackage{fancyhdr}
\usepackage{caption}
\usepackage{subcaption}
\usepackage{xurl}
\usepackage{etoolbox}
\usepackage[section]{placeins}
\definecolor{papergray}{HTML}{555555}
\usepackage[authoryear,round]{natbib}
\setcitestyle{aysep={,},yysep={;},notesep={; }}
\usepackage[hidelinks]{hyperref}
\usepackage{tikz}
\usetikzlibrary{arrows.meta,positioning,shapes.geometric,calc,decorations.pathreplacing,fit,backgrounds}

\hypersetup{
  pdftitle={NoLBERT Calls It: What Managers Will Say, and When the Market Knows It},
  pdfauthor={Kevin Foley, Jonathan Hartadi, Shivesh Prakash, Hino Wirachapong Suwanphibun, Swapnil Vatsal},
  pdfsubject={Hindsight-free text forecasts for prediction markets on earnings-call words},
  pdfkeywords={prediction markets, Kalshi, earnings calls, NoLBERT, look-ahead bias, forecasting, price discovery}
}

\setlist{nosep,leftmargin=1.5em}

\titleformat{\section}[hang]
  {\Large\bfseries\color{black}}{\thesection}{0.7em}{}
\titleformat{\subsection}[hang]
  {\large\bfseries\color{black}}{\thesubsection}{0.65em}{}
\titleformat{\subsubsection}[hang]
  {\normalsize\bfseries\color{black}}{\thesubsubsection}{0.6em}{}
\titleformat{\paragraph}[runin]
  {\normalsize\bfseries\color{black}}{}{0pt}{}[.]
\titlespacing*{\section}{0pt}{2.2ex plus 0.6ex minus 0.2ex}{1.0ex}
\titlespacing*{\subsection}{0pt}{1.8ex plus 0.4ex minus 0.2ex}{0.7ex}
\titlespacing*{\paragraph}{0pt}{1.0ex plus 0.3ex minus 0.2ex}{0.6em}

\definecolor{navyblue}{RGB}{20, 55, 90}
\definecolor{midblue}{RGB}{40, 90, 135}
\definecolor{softblue}{RGB}{236, 243, 250}
\definecolor{accentteal}{RGB}{30, 120, 130}
\definecolor{softteal}{RGB}{235, 246, 247}
\definecolor{accentorange}{RGB}{190, 80, 25}
\definecolor{softorange}{RGB}{253, 244, 238}
\definecolor{darkgray}{RGB}{40, 45, 50}
\definecolor{midgray}{RGB}{100, 110, 120}
\definecolor{lightgray}{RGB}{246, 248, 250}
\definecolor{cardbg}{RGB}{255, 255, 255}
\definecolor{bordergray}{RGB}{205, 205, 205}
\definecolor{nolorange}{HTML}{D35A22}
\definecolor{finblue}{HTML}{2B6CC4}
\definecolor{chronoteal}{HTML}{1A9B7E}
\definecolor{debviolet}{HTML}{7558C6}
\definecolor{inkgray}{HTML}{2A2F35}

\fancypagestyle{paperstyle}{%
  \fancyhf{}
  \fancyhead[L]{\footnotesize\color{papergray}\textsc{NoLBERT Calls It}}
  \fancyhead[R]{\footnotesize\color{papergray}\thepage}
  \renewcommand{\headrulewidth}{0.35pt}
  \renewcommand{\headrule}{\hbox to\headwidth{\color{bordergray}\leaders\hrule height \headrulewidth\hfill}}
  
}
\AtBeginEnvironment{thebibliography}{\small\setlength{\bibsep}{0.45em}}

\newcommand{\topA}[1]{\cellcolor{accentteal!42}\textbf{#1}}
\newcommand{\topB}[1]{\cellcolor{accentteal!24}#1}
\newcommand{\topC}[1]{\cellcolor{accentteal!10}#1}
\newcommand{\shadekey}{\setlength{\fboxsep}{1.2pt}%
  Shading marks the best (\colorbox{accentteal!42}{\strut\textbf{darkest}}),
  second (\colorbox{accentteal!24}{\strut second}) and third
  (\colorbox{accentteal!10}{\strut third}) value of each column}
\newcolumntype{L}[1]{>{\raggedright\arraybackslash}p{#1}}
\newcommand{\tablenotes}[1]{%
  \par\vspace{0.45em}
  \begin{minipage}{\linewidth}\footnotesize\textit{Notes:} #1\end{minipage}}



\newcommand{\todo}[1]{\textcolor{gray}{\textit{[#1]}}}
\newlist{outline}{itemize}{1}
\setlist[outline]{label=\textbullet,leftmargin=1.5em,itemsep=0.2em,topsep=0.4em,
                  before=\color{midgray}\small}

\newcommand{\setV}[2]{\expandafter\gdef\csname val@#1\endcsname{#2}}
\newcommand{\V}[1]{\ifcsname val@#1\endcsname\csname val@#1\endcsname\else\todo{\detokenize{#1}}\fi}

\setV{invMarkets}{9,570}
\setV{invSettled}{7,593}
\setV{invEvents}{691}
\setV{invSeries}{225}
\setV{invVolume}{75.2}
\setV{matchCompanies}{224}
\setV{nCompanies}{225}
\setV{settledCalls}{548}
\setV{matchedCalls}{534}
\setV{agreeRule}{99.6\%}
\setV{kappaRule}{0.991}
\setV{agreeEveryone}{97.5\%}
\setV{callsAll}{235,980}
\setV{callsKalshiFirms}{10,044}
\setV{kalshiFirms}{172}
\setV{nWords}{2,641}
\setV{wordsOne}{2,142}
\setV{wordsSome}{462}
\setV{wordsShared}{37}
\setV{nPassages}{410,919}
\setV{synthTotal}{540,041}
\setV{synthTrain}{435,233}
\setV{synthTrainCalls}{7,938}
\setV{synthTrainSaid}{23\%}
\setV{synthVal}{34,577}
\setV{synthValCalls}{636}
\setV{synthValSaid}{28\%}
\setV{synthTest}{70,231}
\setV{synthTestCalls}{1,298}
\setV{synthTestSaid}{33\%}
\setV{hTrain}{103,401}
\setV{hVal}{23,088}
\setV{hTest}{308,744}
\setV{said2006}{18\%}
\setV{said2024}{28\%}
\setV{said2026}{34\%}
\setV{kTrain}{2,642}
\setV{kTrainCalls}{203}
\setV{kTrainFirms}{111}
\setV{kTrainSaid}{57\%}
\setV{kTest}{4,645}
\setV{kTestCalls}{330}
\setV{kTestFirms}{163}
\setV{kTestSaid}{51\%}
\setV{kTestPriced1}{4,581}
\setV{cov168}{69\%}
\setV{cov96}{84\%}
\setV{cov48}{90\%}
\setV{cov24}{95\%}
\setV{cov1}{99\%}
\setV{wordsPerCallMedian}{13}
\setV{maxPassages}{64}
\setV{pLen}{256}
\setV{qLen}{128}
\setV{hidden}{256}
\setV{epochs}{8}
\setV{patience}{2}
\setV{nBoot}{1,000}
\setV{briBaseSyn}{0.2324}
\setV{briRuleSyn}{0.1529}
\setV{aucRuleSyn}{0.783}
\setV{briLexSyn}{0.1255}
\setV{aucLexSyn}{0.875}
\setV{briNolSyn}{0.1221}
\setV{aucNolSyn}{0.887}
\setV{briTextNolSyn}{0.1702}
\setV{aucTextNolSyn}{0.795}
\setV{briFinSyn}{0.1215}
\setV{aucFinSyn}{0.889}
\setV{briTextFinSyn}{0.1703}
\setV{aucTextFinSyn}{0.796}
\setV{briDebSyn}{0.1220}
\setV{aucDebSyn}{0.887}
\setV{briTextDebSyn}{0.1735}
\setV{aucTextDebSyn}{0.790}
\setV{briChrSyn}{0.1226}
\setV{aucChrSyn}{0.885}
\setV{briTextChrSyn}{0.1730}
\setV{aucTextChrSyn}{0.783}
\setV{briEncRangeSyn}{0.1215--0.1226}
\setV{briBaseK}{0.3232}
\setV{briRuleK}{0.2289}
\setV{aucRuleK}{0.694}
\setV{briLexK}{0.1938}
\setV{aucLexK}{0.783}
\setV{briNolK}{0.1893}
\setV{aucNolK}{0.789}
\setV{briTextNolK}{0.2873}
\setV{aucTextNolK}{0.641}
\setV{briFinK}{0.1888}
\setV{aucFinK}{0.790}
\setV{briTextFinK}{0.2851}
\setV{aucTextFinK}{0.628}
\setV{briDebK}{0.1899}
\setV{aucDebK}{0.787}
\setV{briTextDebK}{0.2905}
\setV{aucTextDebK}{0.630}
\setV{briChrK}{0.1904}
\setV{aucChrK}{0.788}
\setV{briTextChrK}{0.2790}
\setV{aucTextChrK}{0.636}
\setV{briEncRangeK}{0.1888--0.1904}
\setV{briPriceK}{0.1626}
\setV{aucPriceK}{0.842}
\setV{dNolLexSyn}{\textminus{}0.0034}
\setV{dNolLexSynCI}{[\textminus{}0.0038, \textminus{}0.0027]}
\setV{dAucNolLexSyn}{+0.011}
\setV{dFinLexSyn}{\textminus{}0.0040}
\setV{dFinLexSynCI}{[\textminus{}0.0046, \textminus{}0.0032]}
\setV{dAucFinLexSyn}{+0.014}
\setV{dDebLexSyn}{\textminus{}0.0035}
\setV{dDebLexSynCI}{[\textminus{}0.0039, \textminus{}0.0028]}
\setV{dAucDebLexSyn}{+0.011}
\setV{dChrLexSyn}{\textminus{}0.0029}
\setV{dChrLexSynCI}{[\textminus{}0.0034, \textminus{}0.0022]}
\setV{dAucChrLexSyn}{+0.010}
\setV{dNolLexK}{\textminus{}0.0045}
\setV{dNolLexKCI}{[\textminus{}0.0063, \textminus{}0.0025]}
\setV{dAucNolLexK}{+0.006}
\setV{dFinLexK}{\textminus{}0.0051}
\setV{dFinLexKCI}{[\textminus{}0.0075, \textminus{}0.0026]}
\setV{dAucFinLexK}{+0.007}
\setV{dDebLexK}{\textminus{}0.0040}
\setV{dDebLexKCI}{[\textminus{}0.0058, \textminus{}0.0020]}
\setV{dAucDebLexK}{+0.005}
\setV{dChrLexK}{\textminus{}0.0035}
\setV{dChrLexKCI}{[\textminus{}0.0055, \textminus{}0.0015]}
\setV{dAucChrLexK}{+0.005}
\setV{pctNolLexSyn}{2.7\%}
\setV{pctNolLexK}{2.3\%}
\setV{price168}{0.1755}
\setV{rawPrice168}{0.1748}
\setV{nolAlone168}{0.1830}
\setV{nolPlus168}{0.1715}
\setV{spread168}{8.0}
\setV{price96}{0.1742}
\setV{rawPrice96}{0.1739}
\setV{nolAlone96}{0.1860}
\setV{nolPlus96}{0.1720}
\setV{spread96}{7.1}
\setV{price72}{0.1735}
\setV{rawPrice72}{0.1730}
\setV{nolAlone72}{0.1864}
\setV{nolPlus72}{0.1717}
\setV{spread72}{6.6}
\setV{price48}{0.1708}
\setV{rawPrice48}{0.1704}
\setV{nolAlone48}{0.1868}
\setV{nolPlus48}{0.1695}
\setV{spread48}{5.6}
\setV{price24}{0.1695}
\setV{rawPrice24}{0.1697}
\setV{nolAlone24}{0.1870}
\setV{nolPlus24}{0.1684}
\setV{spread24}{4.7}
\setV{price12}{0.1658}
\setV{rawPrice12}{0.1658}
\setV{nolAlone12}{0.1886}
\setV{nolPlus12}{0.1653}
\setV{spread12}{4.3}
\setV{price1}{0.1630}
\setV{rawPrice1}{0.1626}
\setV{nolAlone1}{0.1893}
\setV{nolPlus1}{0.1633}
\setV{spread1}{3.9}
\setV{aucRaw168}{0.818}
\setV{aucNolAlone168}{0.802}
\setV{gainNol168}{0.0040}
\setV{gainNol168CI}{[0.0018, 0.0062]}
\setV{gainNol96}{0.0022}
\setV{gainNol96CI}{[0.0005, 0.0040]}
\setV{gainNol72}{0.0018}
\setV{gainNol72CI}{[0.0003, 0.0034]}
\setV{gainNol48}{0.0013}
\setV{gainNol48CI}{[\textminus{}0.0001, 0.0028]}
\setV{gainNol24}{0.0012}
\setV{gainNol24CI}{[\textminus{}0.0001, 0.0025]}
\setV{gainNol12}{0.0004}
\setV{gainNol12CI}{[\textminus{}0.0003, 0.0011]}
\setV{gainNol1}{\textminus{}0.0003}
\setV{gainNol1CI}{[\textminus{}0.0011, 0.0005]}
\setV{gainPctNol168}{2.3\%}
\setV{gainFin168}{0.0041}
\setV{gainFin168CI}{[0.0020, 0.0062]}
\setV{gainFin96}{0.0025}
\setV{gainFin96CI}{[0.0008, 0.0044]}
\setV{gainFin72}{0.0021}
\setV{gainFin72CI}{[0.0005, 0.0036]}
\setV{gainFin48}{0.0016}
\setV{gainFin48CI}{[0.0001, 0.0031]}
\setV{gainFin24}{0.0014}
\setV{gainFin24CI}{[0.0002, 0.0028]}
\setV{gainFin12}{0.0006}
\setV{gainFin12CI}{[\textminus{}0.0002, 0.0013]}
\setV{gainFin1}{\textminus{}0.0001}
\setV{gainFin1CI}{[\textminus{}0.0009, 0.0007]}
\setV{gainPctFin168}{2.4\%}
\setV{gainDeb168}{0.0036}
\setV{gainDeb168CI}{[0.0014, 0.0058]}
\setV{gainDeb96}{0.0020}
\setV{gainDeb96CI}{[0.0003, 0.0038]}
\setV{gainDeb72}{0.0016}
\setV{gainDeb72CI}{[0.0001, 0.0031]}
\setV{gainDeb48}{0.0011}
\setV{gainDeb48CI}{[\textminus{}0.0003, 0.0025]}
\setV{gainDeb24}{0.0010}
\setV{gainDeb24CI}{[\textminus{}0.0003, 0.0022]}
\setV{gainDeb12}{0.0004}
\setV{gainDeb12CI}{[\textminus{}0.0003, 0.0011]}
\setV{gainDeb1}{\textminus{}0.0004}
\setV{gainDeb1CI}{[\textminus{}0.0011, 0.0005]}
\setV{gainPctDeb168}{2.0\%}
\setV{gainChr168}{0.0037}
\setV{gainChr168CI}{[0.0014, 0.0059]}
\setV{gainChr96}{0.0019}
\setV{gainChr96CI}{[0.0002, 0.0037]}
\setV{gainChr72}{0.0016}
\setV{gainChr72CI}{[0.0001, 0.0031]}
\setV{gainChr48}{0.0010}
\setV{gainChr48CI}{[\textminus{}0.0004, 0.0025]}
\setV{gainChr24}{0.0010}
\setV{gainChr24CI}{[\textminus{}0.0003, 0.0023]}
\setV{gainChr12}{0.0004}
\setV{gainChr12CI}{[\textminus{}0.0003, 0.0011]}
\setV{gainChr1}{\textminus{}0.0004}
\setV{gainChr1CI}{[\textminus{}0.0012, 0.0004]}
\setV{gainPctChr168}{2.1\%}
\setV{gainLex168}{0.0031}
\setV{gainLex168CI}{[0.0010, 0.0052]}
\setV{gainLex96}{0.0015}
\setV{gainLex96CI}{[\textminus{}0.0004, 0.0035]}
\setV{gainLex72}{0.0013}
\setV{gainLex72CI}{[\textminus{}0.0005, 0.0029]}
\setV{slopeOutNol168}{0.32}
\setV{seOutNol168}{0.05}
\setV{slopeMoveNol168}{0.32}
\setV{seMoveNol168}{0.02}
\setV{slopeOutNol96}{0.25}
\setV{seOutNol96}{0.05}
\setV{slopeMoveNol96}{0.24}
\setV{seMoveNol96}{0.02}
\setV{slopeOutNol48}{0.18}
\setV{seOutNol48}{0.05}
\setV{slopeMoveNol48}{0.18}
\setV{seMoveNol48}{0.02}
\setV{slopeOutNol24}{0.18}
\setV{seOutNol24}{0.04}
\setV{slopeMoveNol24}{0.15}
\setV{seMoveNol24}{0.01}
\setV{slopeOutNol12}{0.09}
\setV{seOutNol12}{0.04}
\setV{slopeMoveNol12}{0.10}
\setV{seMoveNol12}{0.01}
\setV{slopeOutLex168}{0.27}
\setV{seOutLex168}{0.05}
\setV{slopeMoveLex168}{0.28}
\setV{seMoveLex168}{0.02}
\setV{slopeOutLex96}{0.21}
\setV{seOutLex96}{0.05}
\setV{slopeMoveLex96}{0.22}
\setV{seMoveLex96}{0.02}
\setV{slopeOutLex48}{0.14}
\setV{seOutLex48}{0.05}
\setV{slopeMoveLex48}{0.17}
\setV{seMoveLex48}{0.02}
\setV{slopeOutLex24}{0.14}
\setV{seOutLex24}{0.04}
\setV{slopeMoveLex24}{0.13}
\setV{seMoveLex24}{0.01}
\setV{slopeOutLex12}{0.06}
\setV{seOutLex12}{0.04}
\setV{slopeMoveLex12}{0.09}
\setV{seMoveLex12}{0.01}
\setV{slopeOutFin168}{0.33}
\setV{seOutFin168}{0.05}
\setV{slopeMoveFin168}{0.31}
\setV{seMoveFin168}{0.02}
\setV{slopeOutFin96}{0.26}
\setV{seOutFin96}{0.05}
\setV{slopeMoveFin96}{0.23}
\setV{seMoveFin96}{0.02}
\setV{slopeOutFin48}{0.19}
\setV{seOutFin48}{0.05}
\setV{slopeMoveFin48}{0.18}
\setV{seMoveFin48}{0.02}
\setV{slopeOutFin24}{0.19}
\setV{seOutFin24}{0.04}
\setV{slopeMoveFin24}{0.15}
\setV{seMoveFin24}{0.01}
\setV{slopeOutFin12}{0.10}
\setV{seOutFin12}{0.04}
\setV{slopeMoveFin12}{0.10}
\setV{seMoveFin12}{0.01}
\setV{rsqNol168}{0.651}
\setV{rsqNol1}{0.668}
\setV{corrNol1}{0.84}
\setV{rsqFin168}{0.658}
\setV{rsqFin1}{0.670}
\setV{corrFin1}{0.85}
\setV{rsqDeb168}{0.663}
\setV{rsqDeb1}{0.667}
\setV{corrDeb1}{0.84}
\setV{rsqChr168}{0.640}
\setV{rsqChr1}{0.662}
\setV{corrChr1}{0.84}
\setV{rsqLex168}{0.609}
\setV{rsqLex1}{0.659}
\setV{corrLex1}{0.84}
\setV{eFinSyn}{\textminus{}0.0006}
\setV{eFinSynCI}{[\textminus{}0.0013, 0.0001]}
\setV{eFinSynCIcalls}{[\textminus{}0.0009, \textminus{}0.0004]}
\setV{eFinSynPairs}{75\%}
\setV{eFinK}{\textminus{}0.0006}
\setV{eFinKCI}{[\textminus{}0.0022, 0.0011]}
\setV{eFinKCIcalls}{[\textminus{}0.0019, 0.0007]}
\setV{eFinKPairs}{66\%}
\setV{eFinVal}{\textminus{}0.0004}
\setV{eFinValCI}{[\textminus{}0.0009, 0.0001]}
\setV{eFinValCIcalls}{[\textminus{}0.0006, \textminus{}0.0001]}
\setV{eFinValPairs}{69\%}
\setV{eFinPlus168}{\textminus{}0.0002}
\setV{eFinPlus168CI}{[\textminus{}0.0007, 0.0004]}
\setV{eFinPlusMaxAbs}{0.0003}
\setV{eDebSyn}{\textminus{}0.0001}
\setV{eDebSynCI}{[\textminus{}0.0008, 0.0005]}
\setV{eDebSynCIcalls}{[\textminus{}0.0003, 0.0001]}
\setV{eDebSynPairs}{52\%}
\setV{eDebK}{0.0005}
\setV{eDebKCI}{[\textminus{}0.0008, 0.0018]}
\setV{eDebKCIcalls}{[\textminus{}0.0004, 0.0015]}
\setV{eDebKPairs}{36\%}
\setV{eDebVal}{0.0002}
\setV{eDebValCI}{[\textminus{}0.0002, 0.0007]}
\setV{eDebValCIcalls}{[0.0000, 0.0005]}
\setV{eDebValPairs}{34\%}
\setV{eDebPlus168}{0.0004}
\setV{eDebPlus168CI}{[\textminus{}0.0001, 0.0009]}
\setV{eDebPlusMaxAbs}{0.0004}
\setV{eChrSyn}{0.0005}
\setV{eChrSynCI}{[\textminus{}0.0002, 0.0011]}
\setV{eChrSynCIcalls}{[0.0003, 0.0007]}
\setV{eChrSynPairs}{42\%}
\setV{eChrK}{0.0010}
\setV{eChrKCI}{[\textminus{}0.0003, 0.0024]}
\setV{eChrKCIcalls}{[\textminus{}0.0002, 0.0022]}
\setV{eChrKPairs}{28\%}
\setV{eChrVal}{0.0002}
\setV{eChrValCI}{[\textminus{}0.0002, 0.0006]}
\setV{eChrValCIcalls}{[0.0000, 0.0005]}
\setV{eChrValPairs}{46\%}
\setV{eChrPlus168}{0.0003}
\setV{eChrPlus168CI}{[\textminus{}0.0002, 0.0008]}
\setV{eChrPlusMaxAbs}{0.0003}
\setV{seedSdSynLo}{0.0007}
\setV{seedSdSynHi}{0.0008}
\setV{finSeedMin}{0.1214}
\setV{finSeedMax}{0.1237}
\setV{nolSeedMin}{0.1218}
\setV{nolSeedMax}{0.1240}
\setV{hNolTwSeen}{0.0876}
\setV{hNolTwUnseen}{0.1279}
\setV{hNolTwK}{0.1971}
\setV{hNolSeen}{0.0879}
\setV{hNolUnseen}{0.1292}
\setV{hNolK}{0.1992}
\setV{hFinSeen}{0.0894}
\setV{hFinUnseen}{0.1350}
\setV{hFinK}{0.2137}
\setV{hDebSeen}{0.0881}
\setV{hDebUnseen}{0.1296}
\setV{hDebK}{0.2000}
\setV{hChrSeen}{0.0879}
\setV{hChrUnseen}{0.1288}
\setV{hChrK}{0.1984}
\setV{hChrTwSeen}{0.0882}
\setV{hChrTwUnseen}{0.1293}
\setV{hChrTwK}{0.2001}
\setV{hLexSeen}{0.0883}
\setV{hLexUnseen}{0.1266}
\setV{premFin}{0.0053}
\setV{premFinCI}{[0.0046, 0.0059]}
\setV{premDeb}{0.0012}
\setV{premDebCI}{[0.0009, 0.0015]}
\setV{premNol}{0.0010}
\setV{premNolCI}{[0.0008, 0.0011]}
\setV{premChr}{0.0006}
\setV{premChrCI}{[0.0003, 0.0009]}
\setV{premRatioFin}{5.5}
\setV{finGap2013}{0.0005}
\setV{finGap2023}{0.0036}
\setV{finGap2025}{0.0058}
\setV{finGap2026}{0.0085}
\setV{minOtherGap}{0.0000}
\setV{lexBeatsNolTwUnseen}{0.0013}
\setV{chrTwVsNolTwSeen}{+0.0006}
\setV{chrTwVsNolTwSeenCI}{[0.0005, 0.0007]}
\setV{chrTwVsNolTwUnseen}{+0.0014}
\setV{chrTwVsNolTwUnseenCI}{[0.0011, 0.0017]}
\setV{chrTwVsNolTwK}{+0.0030}
\setV{chrTwVsNolTwKCI}{[0.0016, 0.0045]}
\setV{catPlace168Gain}{0.0129}
\setV{catPlace168GainCI}{[0.0064, 0.0192]}
\setV{catPlace168N}{218}
\setV{catPlace168Price}{0.209}
\setV{catPlace168Model}{0.187}
\setV{catTech168Gain}{0.0080}
\setV{catTech168GainCI}{[0.0034, 0.0132]}
\setV{catTech168N}{422}
\setV{catTech168Price}{0.176}
\setV{catTech168Model}{0.168}
\setV{catBus168Gain}{0.0068}
\setV{catBus168GainCI}{[0.0034, 0.0103]}
\setV{catBus168N}{763}
\setV{catBus168Price}{0.188}
\setV{catBus168Model}{0.184}
\setV{catMacro168Gain}{0.0053}
\setV{catMacro168GainCI}{[0.0011, 0.0101]}
\setV{catMacro168N}{358}
\setV{catMacro168Price}{0.176}
\setV{catMacro168Model}{0.173}
\setV{catOwn168Gain}{\textminus{}0.0023}
\setV{catOwn168GainCI}{[\textminus{}0.0066, 0.0023]}
\setV{catOwn168N}{503}
\setV{catOwn168Price}{0.156}
\setV{catOwn168Model}{0.188}
\setV{catOtherCo168Gain}{\textminus{}0.0011}
\setV{catOtherCo168GainCI}{[\textminus{}0.0073, 0.0046]}
\setV{catOtherCo168N}{545}
\setV{catOtherCo168Price}{0.169}
\setV{catOtherCo168Model}{0.195}
\setV{catOther168Gain}{0.0036}
\setV{catOther168GainCI}{[\textminus{}0.0019, 0.0090]}
\setV{catOther168N}{367}
\setV{catOther168Price}{0.172}
\setV{catOther168Model}{0.183}
\setV{catAll168Gain}{0.0040}
\setV{catAll168GainCI}{[0.0018, 0.0061]}
\setV{catAll168N}{3,203}
\setV{catAll168Price}{0.175}
\setV{catAll168Model}{0.183}
\setV{catPlace1Gain}{\textminus{}0.0009}
\setV{catPlace1GainCI}{[\textminus{}0.0039, 0.0022]}
\setV{catPlace1N}{330}
\setV{catPlace1Price}{0.184}
\setV{catPlace1Model}{0.201}
\setV{catTech1Gain}{0.0012}
\setV{catTech1GainCI}{[\textminus{}0.0008, 0.0031]}
\setV{catTech1N}{578}
\setV{catTech1Price}{0.171}
\setV{catTech1Model}{0.181}
\setV{catBus1Gain}{0.0022}
\setV{catBus1GainCI}{[0.0008, 0.0036]}
\setV{catBus1N}{1,096}
\setV{catBus1Price}{0.183}
\setV{catBus1Model}{0.188}
\setV{catMacro1Gain}{\textminus{}0.0001}
\setV{catMacro1GainCI}{[\textminus{}0.0019, 0.0017]}
\setV{catMacro1N}{538}
\setV{catMacro1Price}{0.155}
\setV{catMacro1Model}{0.169}
\setV{catOwn1Gain}{\textminus{}0.0022}
\setV{catOwn1GainCI}{[\textminus{}0.0044, \textminus{}0.0001]}
\setV{catOwn1N}{713}
\setV{catOwn1Price}{0.145}
\setV{catOwn1Model}{0.188}
\setV{catOtherCo1Gain}{\textminus{}0.0016}
\setV{catOtherCo1GainCI}{[\textminus{}0.0039, 0.0008]}
\setV{catOtherCo1N}{746}
\setV{catOtherCo1Price}{0.144}
\setV{catOtherCo1Model}{0.196}
\setV{catOther1Gain}{\textminus{}0.0025}
\setV{catOther1GainCI}{[\textminus{}0.0051, 0.0001]}
\setV{catOther1N}{537}
\setV{catOther1Price}{0.162}
\setV{catOther1Model}{0.206}
\setV{catAll1Gain}{\textminus{}0.0003}
\setV{catAll1GainCI}{[\textminus{}0.0011, 0.0005]}
\setV{catAll1N}{4,581}
\setV{catAll1Price}{0.163}
\setV{catAll1Model}{0.189}
\setV{nCatOwn}{796}
\setV{nCatBus}{513}
\setV{nCatOtherCo}{394}
\setV{nCatTech}{326}
\setV{nCatOther}{324}
\setV{nCatMacro}{139}
\setV{nCatPlace}{117}
\setV{nCatPerson}{32}
\setV{trNolCents}{+5.4}
\setV{trNolSe}{1.4}
\setV{trNolT}{3.7}
\setV{trNolN}{1,154}
\setV{trNolEvents}{291}
\setV{trNolDollars}{\$62}
\setV{trNolCapital}{\$667}
\setV{trNolRoc}{9\%}
\setV{trNolHit}{63\%}
\setV{trNolTrainCents}{+8.0}
\setV{trNolTrainT}{4.0}
\setV{trAvgCents}{+3.7}
\setV{trAvgSe}{1.0}
\setV{trAvgT}{3.8}
\setV{trAvgN}{2,277}
\setV{trAvgEvents}{316}
\setV{trAvgDollars}{\$84}
\setV{trAvgCapital}{\$1,374}
\setV{trAvgRoc}{6\%}
\setV{trAvgHit}{64\%}
\setV{trAvgTrainCents}{+5.8}
\setV{trAvgTrainT}{3.9}
\setV{trPriceCents}{+0.3}
\setV{trPriceSe}{1.3}
\setV{trPriceT}{0.2}
\setV{trPriceN}{1,425}
\setV{trPriceEvents}{292}
\setV{trPriceDollars}{\$4}
\setV{trPriceCapital}{\$917}
\setV{trPriceRoc}{0\%}
\setV{trPriceHit}{65\%}
\setV{trPriceTrainCents}{+6.0}
\setV{trPriceTrainT}{2.5}
\setV{trMktCents}{\textminus{}1.0}
\setV{trMktSe}{1.2}
\setV{trMktT}{\textminus{}0.8}
\setV{trMktN}{1,876}
\setV{trMktEvents}{294}
\setV{trMktDollars}{\textminus{}\$19}
\setV{trMktCapital}{\$1,113}
\setV{trMktRoc}{\textminus{}2\%}
\setV{trMktHit}{58\%}
\setV{trMktTrainCents}{+5.1}
\setV{trMktTrainT}{2.5}
\setV{trNolQtwoCents}{+3.5}
\setV{trNolQtwoN}{346}
\setV{trNolQtwoT}{1.3}
\setV{trNolQthreeCents}{+7.0}
\setV{trNolQthreeN}{552}
\setV{trNolQthreeT}{3.4}
\setV{trNolQfourCents}{+4.4}
\setV{trNolQfourN}{256}
\setV{trNolQfourT}{1.4}
\setV{trNolNoCents}{+4.5}
\setV{trNolNoN}{1,070}
\setV{trNolNoT}{3.0}
\setV{trNolYesCents}{+16.4}
\setV{trNolYesN}{84}
\setV{trNolYesT}{3.5}
\setV{trNolHOneCents}{+6.6}
\setV{trNolHOneN}{485}
\setV{trNolHOneT}{3.0}
\setV{trNolRoundCents}{+5.4}
\setV{trNolRoundN}{1,034}
\setV{trNolRoundT}{3.6}
\setV{trOldCents}{+5.1}
\setV{trOldSe}{2.1}
\setV{trOldN}{530}
\setV{cvSelected}{four-encoder average + price}
\setV{cvSelectedScore}{0.14705}
\setV{cvNolScore}{0.14706}
\setV{cvPriceScore}{0.14819}
\setV{cmbAvg168}{0.0042}
\setV{cmbNol168}{0.0043}
\setV{trNolFirstTrainT}{4.0}
\setV{trNolFirstTestCents}{+5.4}
\setV{trNolFirstTestT}{3.7}
\setV{trNolFirstH}{0}
\setV{trNolFirstMargin}{5}
\setV{trNolSettleTrainT}{3.8}
\setV{trNolSettleTestCents}{+3.7}
\setV{trNolSettleTestT}{2.2}
\setV{trNolSettleH}{48}
\setV{trNolSettleMargin}{5}
\setV{trNolExitTrainT}{\textminus{}1.9}
\setV{trNolExitTestCents}{+0.9}
\setV{trNolExitTestT}{0.7}
\setV{trNolExitH}{168}
\setV{trNolExitMargin}{8}
\setV{trPriceBestRule}{one-horizon}
\setV{trPriceBestCents}{\textminus{}0.3}
\setV{trPriceBestT}{\textminus{}0.2}
\setV{trMktBestRule}{one-horizon}
\setV{trMktBestCents}{+1.2}
\setV{trMktBestT}{1.0}

\newcommand{\orcidicon}{%
  \href{https://orcid.org/0009-0007-4120-0921}{%
    \raisebox{-0.8pt}{\includegraphics[height=9pt]{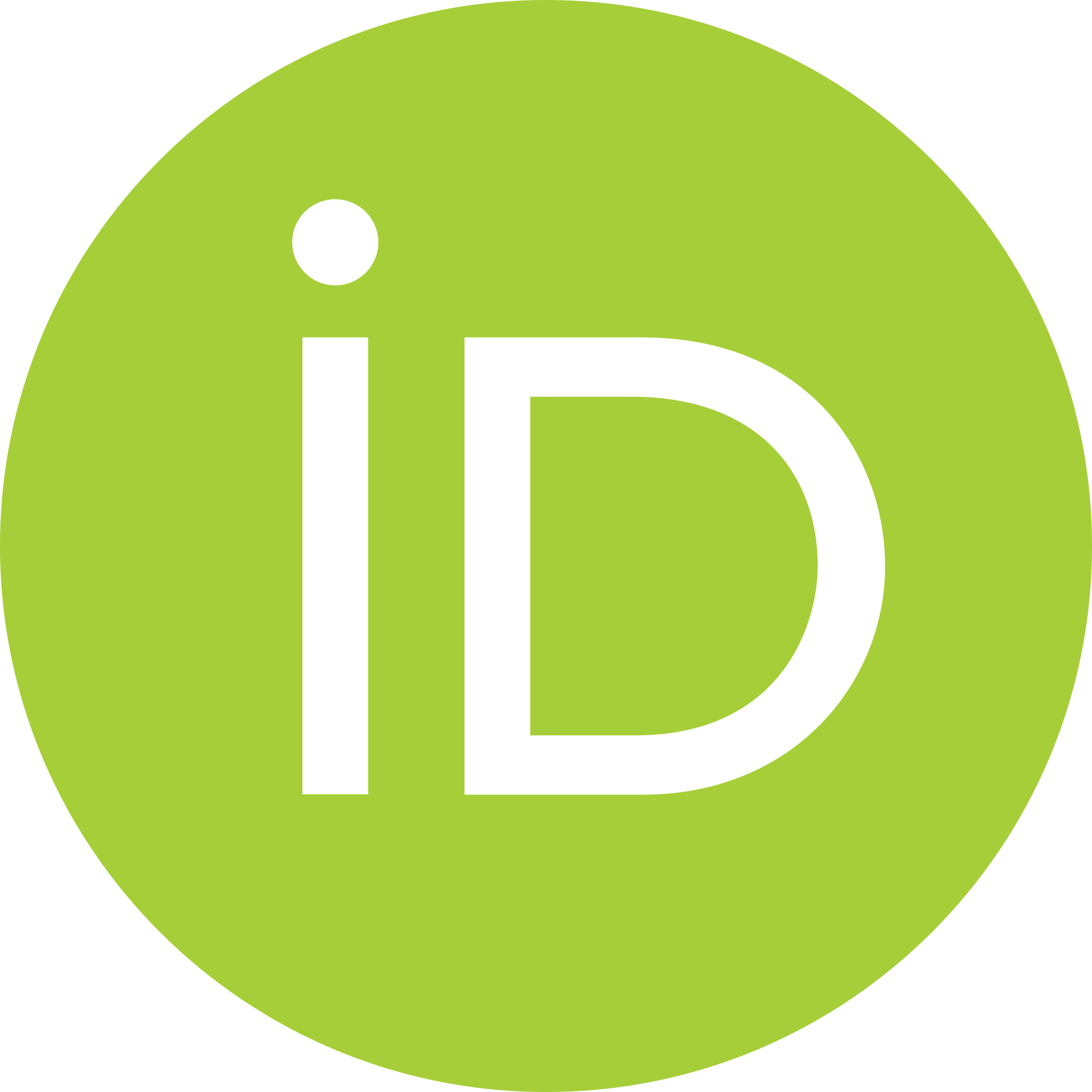}}}}

\renewcommand{\maketitle}{%
  \thispagestyle{empty}
  \vspace*{0.42in}
  \begin{center}
    {\fontsize{24}{29}\selectfont\bfseries
      NoLBERT Calls It:\\[0.12em]
      {\fontsize{19}{24}\selectfont What Managers Will Say, and When the Market Knows It}\par}
    \vspace{1.35em}
    {\large
      Kevin Foley\textsuperscript{*}\quad
      Jonathan Hartadi\textsuperscript{*}\\[0.35em]
      Shivesh Prakash\textsuperscript{*,\(\dagger\)}\,\orcidicon\quad
      Hino Wirachapong Suwanphibun\textsuperscript{*}\quad
      Swapnil Vatsal\textsuperscript{*}\par}
    \vspace{0.75em}
    {\normalsize University of California, Berkeley\par}
    {\small Berkeley, California, USA\par}
    \vspace{0.85em}
    {\small
      \href{mailto:kevinfoley3@berkeley.edu}{kevinfoley3@berkeley.edu}
      \enspace\textperiodcentered\enspace
      \href{mailto:johartadi23@berkeley.edu}{johartadi23@berkeley.edu}\\[0.25em]
      \href{mailto:shivesh_prakash@berkeley.edu}{shivesh\_prakash@berkeley.edu}
      \enspace\textperiodcentered\enspace
      \href{mailto:wirachapong.suw@berkeley.edu}{wirachapong.suw@berkeley.edu}
      \enspace\textperiodcentered\enspace
      \href{mailto:swapnil_vatsal@berkeley.edu}{swapnil\_vatsal@berkeley.edu}\par}
  \end{center}
  \vspace{1.45em}
  \begingroup
    \renewcommand{\thefootnote}{\fnsymbol{footnote}}%
    \footnotetext[1]{Equal contribution; authors listed alphabetically by last name. }%
    \footnotetext[2]{Corresponding author.}%
  \endgroup
}

\usepackage{graphicx}
\usepackage{booktabs}
\usepackage{array}
\usepackage{makecell}
\usepackage{adjustbox}
\usepackage{textcomp}
\makeatletter
\@ifundefined{NC@find@L}{\newcolumntype{L}[1]{>{\raggedright\arraybackslash}p{#1}}}{}
\makeatother

\providecommand{\tablenotes}[1]{\par\smallskip\noindent\parbox{\linewidth}{\footnotesize #1}}

\renewcommand{\textminus}{\ensuremath{-}}
\begin{document}
\maketitle

\begin{abstract}
\noindent
Kalshi sells contracts that pay a dollar if a company's executives say a given word on their next earnings call. We show that a language model with no view of the future can forecast these words, and that a week before the call it knows something the market does not. NoLBERT-Mentions reads a firm's two previous calls through NoLBERT-Calls, an encoder pretrained on text from 1976 to 1995 and adapted to calls up to 2023, and adds the word's mention history. On \V{synthTest} contracts from 2025--26 calls it improves on a history-only model by \V{pctNolLexSyn} in Brier score, level with encoders pretrained on modern financial and web text. Added to the Kalshi price a week before the call, it cuts the price's Brier score by \V{gainPctNol168}; the price then moves a third of the way toward the model, as far as the outcomes justify, and has caught up by the hour of the call. Trading on the disagreement earns \V{trNolCents} cents per contract after fees on later calls ($t=\V{trNolT}$), against \V{trPriceCents} cents for the same rule on the price alone. Backtests on the years an encoder has read flatter FinBERT-Calls \V{premRatioFin} times as much as NoLBERT-Calls.

\end{abstract}

\vspace{0.8em}
\begin{center}
\par\smallskip
\textbf{Code and model:} \url{https://github.com/Shivesh777/NoLBERT-Mentions}
\end{center}
\vfill
\clearpage
\pagestyle{paperstyle}

\section{Introduction}
\label{sec:intro}

Kalshi, the first federally licensed prediction market in the United States \citep{burgi2025kalshi}, lists contracts on what companies will say. Before an earnings call it offers a set of words, such as ``Tariff'' or ``AI / Artificial Intelligence'', and each word is a contract that pays one dollar if someone speaking for the company says it during the call. The price is the market's probability that the word will be said. By 18 November 2026 Kalshi had listed \V{invMarkets} of these contracts on \V{invEvents} calls of \V{invSeries} companies, traders had exchanged \V{invVolume} million of them, and Kalshi had gathered them, together with its markets on companies' reported metrics, into a hub for public companies \citep{kalshi2026hub}.

For a language model, these markets come close to an ideal test. The outcome is a fact about language, settled by the transcript, and much of what predicts it is text a model can read: what the firm discussed last quarter, and how its managers talk about it. The benchmark is demanding. Prediction-market prices are good forecasts \citep{wolfers2004prediction,berg2008accuracy}, if less so far from expiry \citep{page2013calibrated}, and Kalshi quotes a price at every horizon from a week before the call to the last hour. A model can be scored against that price at each horizon, and combined with it, to see whether the text tells the market anything it does not already know.

The test means little if the model has read the future. A model trained on text from the years of a backtest can carry knowledge of what happened \citep{sarkar2024lookahead,lopezlira2025memorization}, and for earnings calls the risk is concrete: FinBERT, the most widely used finance encoder, was pretrained on calls from 2004 to 2019 \citep{huang2023finbert}. Kalshi's contracts are recent (the earliest we observe settled in February 2025), so they lie after the training text of every encoder we use. A mention model, however, has to learn from the long record of earlier calls, and anyone who builds one will want to backtest it there. We therefore measure hindsight in the backtest as well as in the market test.

Our model, \emph{NoLBERT-Mentions}, forecasts a contract on word $w$ at call $t$ from what the firm's executives said on calls $t-1$ and $t-2$. A frozen encoder embeds up to \V{maxPassages} passages of each call, the question ``What can you tell us about $w$?'' attends to them, nine features of the word's history at the firm enter alongside, and the model outputs a distribution for the number of mentions, from which the contract's probability follows. The encoder is NoLBERT-Calls-2023 \citep{foley2026nolbertcallsreadingearningscalls,prakash2026nolbertcalls}: NoLBERT \citep{kakhbod2025nolbert}, pretrained only on text from 1976 to 1995, adapted to earnings calls up to 2023. Real contracts are too few to train on, so we build \V{synthTotal} synthetic ones, pairing every call of a Kalshi company since 2006 with the words Kalshi lists for that company and labelling each from the Capital IQ transcript under Kalshi's settlement rule; the labels reproduce Kalshi's settlements on \V{agreeRule} of real contracts. We compare NoLBERT-Calls with FinBERT-, DeBERTa- and ChronoBERT-Calls, adapted to the same calls, on \V{synthTest} synthetic contracts from 2025--26 calls and on \V{kTest} Kalshi contracts from April to November 2026, alone and with the Kalshi price at ten horizons.

The text carries real information about what managers will say. NoLBERT-Mentions improves on the word's history by \V{pctNolLexSyn} on the synthetic test (a Brier difference of \V{dNolLexSyn}, 95\% interval \V{dNolLexSynCI}) and by \V{pctNolLexK} on Kalshi's contracts. Across ten seeds per encoder, NoLBERT-Calls is statistically indistinguishable from the most accurate of the four, although its pretraining text ends in 1995.

On the day of the call the market is the better forecaster: an hour before it, the Kalshi price has a Brier score of \V{briPriceK}, against \V{briNolK} for NoLBERT-Mentions. A week out, however, the price has not yet absorbed what is in the firm's previous calls. Combined with the price, NoLBERT-Mentions lowers its Brier score by \V{gainNol168} \V{gainNol168CI}, or \V{gainPctNol168}. The market then learns. Where the model and the week-ahead price disagree, the price closes \V{slopeMoveNol168} of the gap by the hour of the call, which is the share the outcomes justify (\V{slopeOutNol168}), and the text's contribution falls to zero. The gains come from words about places, technology, business and the economy; for the firm's own products and for other companies, the market is better informed.

The forecast is worth money after fees. A rule that buys the side the combined forecast favours, once its expected profit after Kalshi's fee clears a margin set on earlier calls, earns \V{trNolCents} cents per contract on \V{trNolN} test contracts ($t=\V{trNolT}$), and every quarter of the test period is profitable. The same rule on the price alone earns \V{trPriceCents} cents.

Backtests on the years an encoder has read flatter it, some far more than others. We train the model on calls before 2012 and compare each encoder's lead over an encoder whose text ends in 2012, first on the calls of 2013--23, which the 2023 encoders read, and then on those of 2025--26, which they did not. The difference, a look-ahead premium, is \V{premFin} for FinBERT-Calls, \V{premRatioFin} times NoLBERT-Calls' \V{premNol}. NoLBERT-Calls-2012, which has read nothing written after 2012, is the best or tied encoder in every test year from 2013 to 2026.

These results add to what is known about how prediction markets aggregate information. Kalshi's prices are accurate on the day of the call but slow to absorb public text a week before it, in line with the weaker calibration far from expiry that \citet{page2013calibrated} document, and here the missing information can be named and the speed at which it is priced measured. The results also bear on the use of language models in finance research. Encoders adapted to the same calls forecast equally well, so the choice among them should turn on whether their backtests can be trusted, and NoLBERT's can; the reader that best matches analysts' questions to managers' answers also anticipates what managers will say, before the market does.

We release the code, the trained model and the aggregate results behind every table and figure.\footnote{\url{https://github.com/Shivesh777/NoLBERT-Mentions}. The transcripts are licensed and are not included. The encoder, NoLBERT-Calls-2023, is at \url{https://github.com/Shivesh777/NoLBERT-Calls}.} Section~\ref{sec:related} reviews related work, Section~\ref{sec:data} describes the data, Section~\ref{sec:methods} the method and Section~\ref{sec:results} the results; Section~\ref{sec:discussion} concludes.

\section{Related work}
\label{sec:related}

\paragraph{Prediction markets}
Market prices are typically accurate forecasts that beat moderately sophisticated benchmarks \citep{wolfers2004prediction}, although with heterogeneous beliefs a price pins down the traders' average belief only within bounds \citep{manski2006interpreting}. Election markets beat polls even months before the vote \citep{berg2008accuracy}. Across many markets, prices are well calibrated close to expiry and biased for events further away \citep{page2013calibrated}, and the favourite--longshot bias reflects misperceived probabilities \citep{snowberg2010favorite}. On Kalshi, \citet{burgi2025kalshi} find prices relatively accurate, with a favourite--longshot pattern that differs between traders who post offers and those who take them. The paper closest to ours is \citet{kim2026mention}, who prompt a commercial large language model with news, earlier transcripts and the market price to forecast Kalshi's earnings-call mention markets (856 contracts from 70 calls in 2025), and find that blending the model's forecast with the market's beats the market. We find the same a week before the call, when their forecasts are made, and go further in four directions: a test five times larger, the horizon at which the market catches up, a trained model whose encoders can be backtested over two decades of calls without hindsight, and a measure of how much such backtests flatter each encoder.

\paragraph{Combining forecasts}
Whether the text adds to the price is a question of forecast combination \citep{bates1969combination}, which we pose in the regression form of \citet{fair1990comparing}: if the price contains all the information in the model's forecast, the model gets no weight. When the forecasters are machine-learning models trained on the same data, the spread of their forecasts behaves like disagreement among investors and predicts stock returns \citep{bali2023mfd}. We average over such variation instead: our forecast is the mean of ten heads that differ only in their seed. We score probabilities with the Brier score \citep{brier1950verification}, a strictly proper scoring rule \citep{gneiting2007proper}.

\paragraph{Text as data}
Text is now a standard source of measurement in economics and finance \citep{gentzkow2019text,loughran2011liability}. Patent claims separate intangibles that improve a firm's own processes from those built into its products \citep{chen2023process}, the similarity between a firm's 10-K and other firms' new patents measures its exposure to creative destruction \citep{kakhbod2024creative}, and a causal text model shows that a patent with a female lead inventor receives fewer citations than it would with a male lead inventor \citep{hochberg2026patents}. Narrative factors estimated from the \emph{Wall Street Journal} price the cross-section of stock returns \citep{bybee2023narrative}, as do topic models of financial news built from word counts or from contextual sentence embeddings \citep{foley2026wordcountscontexttopic,prakash2026wordcounts}. From FOMC records, \citet{kakhbod2026fedmind} measure whether the Fed attributes inflation to demand or to supply, and find that since Volcker it has responded more aggressively to inflation it sees as demand-driven. Delivery matters as well: markets respond to the tone of the Fed Chair's voice \citep{gorodnichenko2023voice} and to the Chair's facial expressions \citep{curti2023face}.

\paragraph{What firms say on calls}
Counts of words and phrases in earnings calls are the raw material of firm-level measures of political risk \citep{hassan2019firmlevel} and climate exposure \citep{sautner2023climate}, and the discussion with analysts on a call carries information beyond the prepared remarks \citep{matsumoto2011calls}. This work, like the measures above, takes what was said as given. Kalshi's markets make whether a word will be said the object of a forecast, so they let us ask how predictable such counts are, and from what.

\paragraph{Machine learning in finance}
Deep networks estimate the stochastic discount factor from a large set of firm and macroeconomic variables \citep{chen2024deep} and model mortgage delinquency, foreclosure and prepayment \citep{sadhwani2021mortgage}. Inverse reinforcement learning recovers the objective behind firms' investment and financing decisions and finds that managers underestimate the shadow cost of capital \citep{nikolov2025purpose}. Interpretable models need not give up accuracy. Glass-box models of corporate bond returns match state-of-the-art accuracy with a structure that shows the nonlinear effects and interactions behind their forecasts \citep{bell2026glassbox}, and AlphaGlass optimizes long-short equity portfolios whose weights trace back to individual characteristics \citep{bell2026alphaglass}. Panel trees split the cross-section of returns into transparent test assets \citep{cong2025ptree}, and an interpretable mortgage-scoring model that builds in a fairness constraint narrows the shortfall for minority borrowers \citep{bell2025fairness}. Embeddings, the representation that language models give to unstructured data, now enter economic models as well: portfolio holdings yield embeddings of assets \citep{gabaix2025embeddings}, and embeddings of product descriptions and images identify substitutes and price elasticities in demand estimation \citep{compiani2026demand,bach2025adventures}. NoLBERT-Mentions borrows from both lines. The passages enter as frozen embeddings, and the head on top is small: besides its attention over the passages, it reads seven named statistics of the question--passage similarities and nine features of the word's history, and a variant without the history shows what the text knows by itself.

\paragraph{Hindsight in language models}
Pretrained language models can carry knowledge of the future into a historical test \citep{sarkar2024lookahead} and recall economic data from their training period \citep{lopezlira2025memorization}; \citet{glasserman2023lookahead} and \citet{levy2024caution} document the consequences in finance and accounting. Asked to stand in for survey respondents, large language models reproduce how investment preferences differ with income, gender and age, but by default answer like young, high-income investors \citep{fedyk2026ai}; \citet{ludwig2025llm} set out when such outputs can be used in econometric work. Point-in-time models remove look-ahead by construction: StoriesLM \citep{sarkar2024storieslm}, ChronoBERT \citep{he2025chronologically}, DatedGPT \citep{yan2026datedgpt} and NoLBERT \citep{kakhbod2025nolbert}, which is pretrained only on text from 1976 to 1995. NoLBERT has since been used to build topics from financial news for the cross-section of stock returns \citep{foley2026readingnewshindsightnolbert,prakash2026hindsight} and adapted to earnings calls as NoLBERT-Calls \citep{foley2026nolbertcallsreadingearningscalls,prakash2026nolbertcalls}. NoLBERT-Calls matches analysts' questions to managers' answers better than encoders adapted to the same calls from FinBERT, DeBERTa or ChronoBERT, and having read the test years is worth less than a point of retrieval accuracy to any of them. Mention markets let us measure the same premium in a forecasting task, against a market.

\Needspace{6\baselineskip}
\section{Data}
\label{sec:data}

\subsection{Kalshi's mention markets}
Each of Kalshi's earnings-mention series covers one company (``What will Apple say during their next earnings call?''). Before each call Kalshi lists a set of words, a median of \V{wordsPerCallMedian} per call, and each word is a binary contract that pays \$1 if someone speaking for the company says the word, or a form that Kalshi's rules accept, during the call. Most contracts require one mention; some require two, three or five (``Tariff (3+ times)''). The rules count everyone speaking for the company, including the operator, but not the analysts. We pulled every earnings-mention market from Kalshi's public API on 18 November 2026: \V{invMarkets} markets on \V{invEvents} calls of \V{invSeries} companies, of which \V{invSettled} had settled. Over all calls Kalshi has listed \V{nWords} distinct words: \V{wordsOne} for a single company, \V{wordsSome} for two to nine companies, and \V{wordsShared} for ten or more; we call these shared words, and the most widely listed are Tariff, Headwind, Tailwind and AI.

For every contract and each horizon $h \in \{168, 96, 72, 48, 24, 12, 6, 3, 2, 1\}$ hours before the call starts, the price $p_h$ is the mid-quote of the last hourly candle that ended at least $h$ hours before the call, or the last trade if the quoted spread is wider than 25 cents. We also keep the bid and the ask at $h$. Not every contract is quoted a week ahead: on the test calls, \V{cov168} of contracts have a price at 168 hours and \V{cov1} at one hour.

\subsection{Transcripts and labels}
Transcripts come from S\&P Capital IQ through WRDS \citep{ciqtranscripts}: \V{callsAll} calls from 2006 to November 2026, nearly all of US companies, of which \V{callsKalshiFirms} belong to the \V{kalshiFirms} Kalshi companies that have transcripts. We match \V{matchCompanies} of the \V{nCompanies} Kalshi companies to Capital IQ and \V{matchedCalls} of the \V{settledCalls} settled calls to their transcripts. A contract is labelled from the transcript by counting, under Kalshi's rules, how often the company's speakers, including the operator, say the word or an accepted form. These labels agree with Kalshi's settlements on \V{agreeRule} of the contracts (Cohen's $\kappa = \V{kappaRule}$); counting the analysts as well would lower the agreement to \V{agreeEveryone}. Every model is scored against Kalshi's own settlements.

\subsection{Synthetic contracts}
Kalshi's markets began too recently to train a model on, so we build synthetic contracts from history. Every call of a Kalshi company from 2006 on is paired with each word Kalshi has listed for that company and with the \V{wordsShared} shared words, and labelled as above: \V{synthTotal} contracts in all. Each contract carries nine features of the word's history at the firm, computed from earlier calls only: whether it was said on the last call and how often, the shares of the last four and the last eight calls that said it, whether any of the last eight did, whether analysts said it on the last call, the share of all firms' calls in the previous 30 days that said it, the contract's threshold, and the length of the firm's history. The share of synthetic contracts that would have settled Yes rises from \V{said2006} for 2006 calls to \V{said2026} for 2026 calls (Figure~\ref{fig:data}a), consistent with Kalshi listing words about current topics, which executives used less often in earlier years.

\begin{table}[tbp]
\centering
\small
\caption{Samples. A contract is one word on one call. Synthetic contracts pair every call of a Kalshi company with the words Kalshi has listed for that company and the shared words, and are labelled from the transcript; Kalshi contracts are the settled markets, labelled by Kalshi's settlement. ``Said'' is the share that settle Yes.}
\label{tab:samples}
\begin{tabular}{@{}lllrr@{}}
\toprule
Sample & Calls held & Calls & Contracts & Said (\%) \\
\midrule
\multicolumn{5}{l}{\textit{Synthetic, main protocol}} \\
\quad Train & 2006--2023 & 7,938 & 435,233 & 23.4 \\
\quad Validation & 2024 & 636 & 34,577 & 28.5 \\
\quad Test & 2025--2026 & 1,298 & 70,231 & 33.4 \\
\addlinespace\multicolumn{5}{l}{\textit{Synthetic, pre-2012 protocol}} \\
\quad Train & 2006--2011 & 1,856 & 103,401 & 19.5 \\
\quad Validation & 2012 & 422 & 23,088 & 20.2 \\
\quad Test, years the 2023 encoders read & 2013--2023 & 5,660 & 308,744 & 24.9 \\
\quad Test, years no encoder read & 2025--2026 & 1,298 & 70,231 & 33.4 \\
\addlinespace\multicolumn{5}{l}{\textit{Kalshi mention markets}} \\
\quad Train (combiner, trading margin) & Feb 2025--Mar 2026 & 203 & 2,642 & 56.9 \\
\quad Test & Apr--Nov 2026 & 330 & 4,645 & 50.5 \\
\bottomrule
\end{tabular}
 
\tablenotes{The 2023 encoders were adapted on calls up to 2023, so they have read the calls of the pre-2012 protocol's 2013--2023 test (Section~\ref{sec:results:hindsight}). No encoder has read a call from 2025--26. The Kalshi train contracts are used only to fit the combination of model and price and to choose the trading rule's margin; no model is trained on them.}
\end{table}

\subsection{Samples}
\begin{figure}[tbp]
\centering
\includegraphics[width=\linewidth]{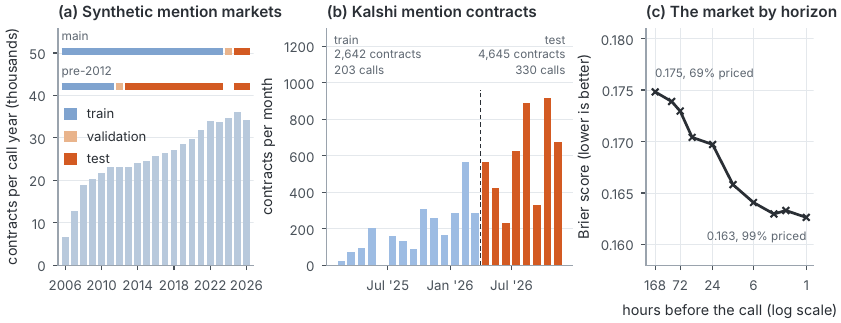}
\caption{The data. (a) Synthetic contracts per call year, with the train, validation and test years of the two protocols. (b) Kalshi contracts by month of the call; the dashed line is the split between Kalshi train and test. (c) Brier score of the Kalshi price by hours before the call, on the test contracts priced at each horizon; the labels give the score and the share of test contracts priced at 168 and 1 hours.}
\label{fig:data}
\end{figure}

Table~\ref{tab:samples} lists the samples. In the main protocol, models are trained on calls up to 2023, tuned on calls from 2024, and tested on calls from 2025--26, which no encoder has read. In the pre-2012 protocol, models are trained on calls up to 2011, tuned on calls from 2012, and tested both on calls from 2013--23, which the 2023 encoders read during adaptation, and on calls from 2025--26. The Kalshi contracts are split by date at 1 April 2026. The \V{kTrain} contracts on \V{kTrainCalls} calls before that date form the Kalshi train set, used only to fit the combination with the price and to choose the trading rule's margin; the \V{kTest} contracts on \V{kTestCalls} calls of \V{kTestFirms} companies from April to November 2026 form the test set (Figure~\ref{fig:data}b). About half of the Kalshi test contracts settle Yes (\V{kTestSaid}), against a third of the synthetic ones (\V{synthTestSaid}), because a synthetic contract pairs every call with every word ever listed for the company, while Kalshi chooses words for a particular call.

\section{Method}
\label{sec:methods}

\subsection{The forecasting problem}
A contract $i$ asks whether word $w_i$ will be said at least $k_i$ times on call $t_i$ of firm $f_i$, and $y_i \in \{0,1\}$ is the answer: Kalshi's settlement for a real contract, our transcript label for a synthetic one. A forecast made $h$ hours before the call may use the text of the firm's earlier calls, the word's history at the firm and, for real contracts, the Kalshi price $p_{i,h}$. We score a probability forecast $q$ by its Brier score $N^{-1}\sum_i (q_i - y_i)^2$ \citep{brier1950verification}, which is strictly proper \citep{gneiting2007proper}, and also report the area under the ROC curve (AUC), which measures ranking alone.

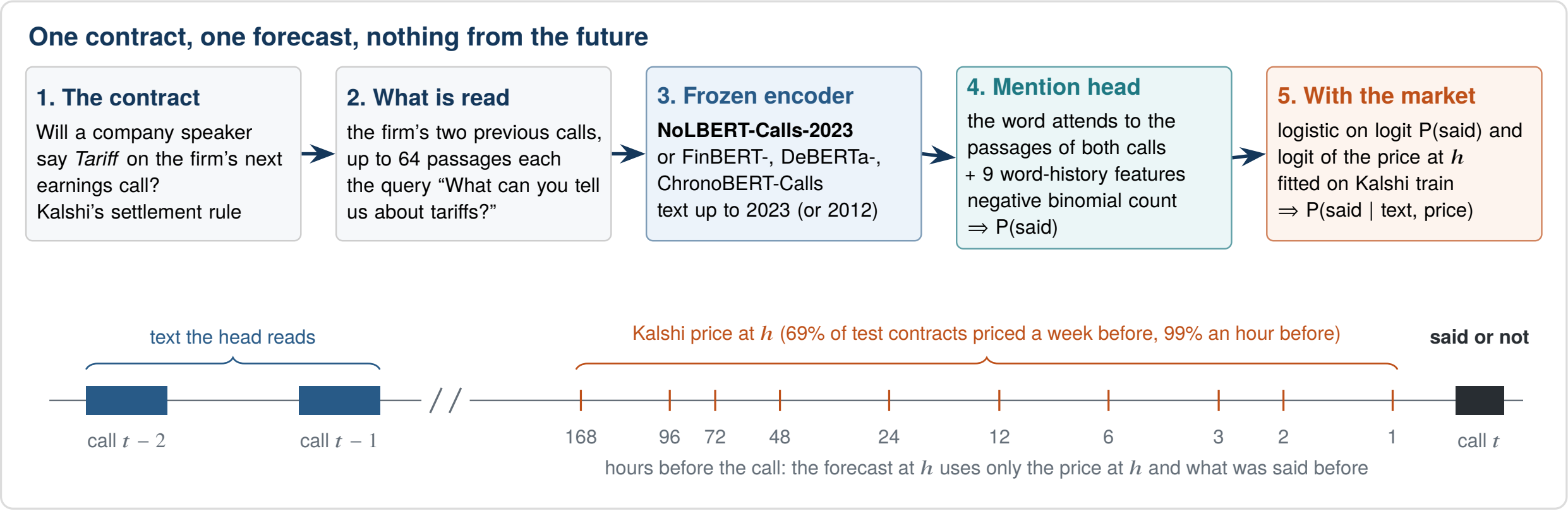
\begin{figure}[tbp]
\centering
\resizebox{\linewidth}{!}{
\begin{tikzpicture}[
    >=Stealth,
    font=\sffamily,
    every node/.append style={execute at begin node={\hyphenpenalty=10000\exhyphenpenalty=10000}},
    panel/.style={draw=bordergray!70, fill=white, rounded corners=3.5pt, line width=0.6pt},
    stepbox/.style={draw=bordergray, fill=lightgray, rounded corners=2pt, line width=0.45pt, text width=2.62cm,
                 minimum height=1.8cm, align=left, inner sep=3.2pt, anchor=north west,
                 font=\sffamily\fontsize{6.2}{7.8}\selectfont},
    annot/.style={font=\sffamily\fontsize{6.1}{7.3}\selectfont\color{midgray}},
    dataflow/.style={->, line width=1.0pt, color=navyblue}
]
\draw[panel] (0, 0.5) rectangle (16.2, 5.75);
\node[anchor=north west, font=\sffamily\bfseries\scriptsize\color{navyblue}] at (0.15, 5.62)
    {One contract, one forecast, nothing from the future};

\node[stepbox] (s1) at (0.25, 5.08)
    {{\bfseries\fontsize{7.2}{8.4}\selectfont\color{navyblue}1. The contract}\\[2pt]
     Will a company speaker say \emph{Tariff} on the firm's next earnings call?\\ Kalshi's settlement rule};
\node[stepbox] (s2) at (3.46, 5.08)
    {{\bfseries\fontsize{7.2}{8.4}\selectfont\color{navyblue}2. What is read}\\[2pt]
     the firm's two previous calls, up to 64 passages each\\
     the query ``What can you tell us about tariffs?''};
\node[stepbox, fill=softblue, draw=midblue!60] (s3) at (6.67, 5.08)
    {{\bfseries\fontsize{7.2}{8.4}\selectfont\color{midblue}3. Frozen encoder}\\[2pt]
     \textbf{NoLBERT-Calls-2023}\\ or FinBERT-, DeBERTa-, ChronoBERT-Calls\\ text up to 2023 (or 2012)};
\node[stepbox, fill=softteal, draw=accentteal!70] (s4) at (9.88, 5.08)
    {{\bfseries\fontsize{7.2}{8.4}\selectfont\color{accentteal}4. Mention head}\\[2pt]
     the word attends to the passages of both calls\\ + 9 word-history features\\ negative binomial count\\ $\Rightarrow$ P(said)};
\node[stepbox, fill=softorange, draw=accentorange!70] (s5) at (13.09, 5.08)
    {{\bfseries\fontsize{7.2}{8.4}\selectfont\color{accentorange}5. With the market}\\[2pt]
     logistic on logit P(said) and logit of the price at $h$\\ fitted on Kalshi train\\ $\Rightarrow$ P(said $\mid$ text, price)};
\foreach \a/\b in {s1/s2, s2/s3, s3/s4, s4/s5}
    \draw[dataflow] (\a.east) -- (\b.west);

\begin{scope}[shift={(0, 1.62)}]
  \draw[midgray, line width=0.45pt] (0.5, 0) -- (4.35, 0);
  \draw[midgray, line width=0.45pt] (4.85, 0) -- (15.75, 0);
  \draw[midgray, line width=0.6pt] (4.44, -0.13) -- (4.56, 0.13);
  \draw[midgray, line width=0.6pt] (4.64, -0.13) -- (4.76, 0.13);
  \foreach \x/\l in {1.3/{call $t-2$}, 3.5/{call $t-1$}} {
    \fill[midblue] (\x-0.42, -0.15) rectangle (\x+0.42, 0.15);
    \node[annot, anchor=north] at (\x, -0.2) {\l};
  }
  \draw[decorate, decoration={brace, amplitude=3.5pt}, midblue, line width=0.5pt] (0.88, 0.32) -- (3.92, 0.32);
  \node[font=\sffamily\fontsize{6.1}{7.3}\selectfont, color=midblue, anchor=south] at (2.4, 0.45) {text the head reads};
  \foreach \h/\x in {168/6.00, 96/6.92, 72/7.39, 48/8.06, 24/9.19, 12/10.33, 6/11.46, 3/12.60, 2/13.27, 1/14.40} {
    \draw[accentorange, line width=0.6pt] (\x, -0.11) -- (\x, 0.11);
    \node[annot, anchor=north] at (\x, -0.16) {\h};
  }
  \node[annot, anchor=north] at (10.2, -0.48) {hours before the call: the forecast at $h$ uses only the price at $h$ and what was said before};
  \draw[decorate, decoration={brace, amplitude=3.5pt}, accentorange, line width=0.5pt] (5.95, 0.32) -- (14.45, 0.32);
  \node[font=\sffamily\fontsize{6.1}{7.3}\selectfont, color=accentorange, anchor=south] at (10.2, 0.45) {Kalshi price at $h$ (\V{cov168} of test contracts priced a week before, \V{cov1} an hour before)};
  \fill[darkgray] (15.05, -0.15) rectangle (15.55, 0.15);
  \node[annot, anchor=north, align=center] at (15.3, -0.2) {call $t$};
  \node[font=\sffamily\bfseries\fontsize{6.1}{7.3}\selectfont, color=darkgray, anchor=south] at (15.3, 0.45) {said or not};
\end{scope}
\end{tikzpicture}}
\caption{How NoLBERT-Mentions forecasts one contract. The head reads what the firm's executives said on its two previous calls through a frozen encoder, together with the word's history at the firm, and outputs a distribution for the number of mentions. With the Kalshi price, a logistic regression fitted on Kalshi train contracts combines the two. A forecast at $h$ hours before the call uses only the price at $h$ and text from earlier calls.}
\label{fig:setup}
\end{figure}

\subsection{NoLBERT-Mentions}
Figure~\ref{fig:setup} summarizes the model. The executives' speech on calls $t-1$ and $t-2$ is cut into passages of 180 words (at most \V{maxPassages} per call, evenly spaced when a call is longer), and each passage is embedded by a frozen encoder with mean pooling over up to \V{pLen} tokens and scaled to unit length. The word enters as a question that lists Kalshi's accepted forms, such as ``What can you tell us about AI or Artificial Intelligence?''. The head projects the question and the passages to \V{hidden} dimensions, adds to each passage an embedding of its type (call $t-1$ or $t-2$, prepared remarks or answers to analysts, and a null slot for firms without earlier calls), and lets the question attend to the passages by scaled dot-product attention. A two-layer perceptron (\V{hidden} units, GELU, dropout 0.1) then reads the question, the attended summary and their product; seven statistics of the raw cosine similarities between question and passages (the maximum, the mean of the top three, the mean and the standard deviation, the maximum within each call, and whether any passage exists); and the nine word-history features. It outputs the mean $\mu$ and dispersion $\alpha$ of a negative binomial distribution for the number of mentions $N$, so that the contract's probability is $q = \Pr(N \ge k)$ and contracts with different thresholds share one model. The loss is the negative log likelihood of the count (capped at 50) plus the cross-entropy of the event. Only the head, about 0.6 million parameters, is trained: with AdamW on synthetic train contracts, 32 calls per batch, for at most \V{epochs} epochs with early stopping on the validation loss, and with the learning rate chosen from two values on the validation set (Appendix Table~\ref{tab:hyper}). Each run uses three seeds; the main runs of the four 2023 encoders use ten. A forecast is the average of the seeds' probabilities.

We call the model with NoLBERT-Calls-2023 \emph{NoLBERT-Mentions}; the same head with another encoder is named after that encoder. We also train the head without the word-history features, to measure what the text knows by itself.

\subsection{Encoders}
All encoders come from NoLBERT-Calls \citep{foley2026nolbertcallsreadingearningscalls,prakash2026nolbertcalls}. NoLBERT-Calls-2023 starts from NoLBERT \citep{kakhbod2025nolbert}, a DeBERTa-v3-base encoder pretrained only on text from 1976 to 1995, and adapts it in two stages, first on calls from all countries up to 2012 (480 minutes) and then on US calls up to 2023 (two runs of 150 minutes), with objectives that match analysts' questions to managers' answers within a call, match passages of the same call and fill in masked words. FinBERT-Calls-2023 \citep[from][]{huang2023finbert}, DeBERTa-Calls-2023 \citep[from][]{he2021debertav3} and ChronoBERT-Calls-2023 \citep[from ChronoBERT-2023,][]{he2025chronologically} were adapted to US calls up to 2023 with one 150-minute run of the same recipe. For the pre-2012 protocol we add two encoders whose text ends in 2012: NoLBERT-Calls-2012, the first stage of NoLBERT-Calls-2023, and ChronoBERT-Calls-2012, which is ChronoBERT-2012 adapted to US calls up to 2012 with the 150-minute recipe. Two of the 2023 encoders have also read text that is not dated: FinBERT's pretraining corpus includes earnings calls from 2004 to 2019, and DeBERTa's web text has no date control.

\paragraph{Baselines} The base rate is the share of synthetic train contracts said. The rule ``said on the last call'' predicts the share said among training contracts with the same value of that feature. The word-history model is a logistic regression on the nine features, trained on synthetic train contracts; it is the bar the text has to clear.

\subsection{Adding the price}
To ask whether a model adds to the market, we fit for each horizon $h$, on the Kalshi train contracts priced at $h$, a logistic regression of the outcome on $\operatorname{logit} p_h$ and $\operatorname{logit} q$, and apply it to the test contracts priced at $h$. This is the regression of \citet{fair1990comparing} in probability form: if the price contained all the information in $q$, the weight on $q$ would be zero. The reference is the price recalibrated in the same way, a logistic regression on $\operatorname{logit} p_h$ alone, whose Brier score is almost identical to the raw price's. The heads never see a Kalshi contract, so their forecasts on Kalshi train contracts are out of sample. Appendix~\ref{app:combiners} compares richer combiners, chosen by cross-validation on Kalshi train.

\paragraph{Price discovery} If the gain from the text fades as the call approaches, the market may be learning what the text knew. For an early horizon $e$, we regress the price's later move, $p_{1} - p_{e}$, and the outcome's surprise, $y - p_{e}$, on the model's disagreement with the early price, $q - p_{e}$, by least squares on the test contracts priced at both $e$ and one hour, with standard errors clustered by call. A positive slope in the first regression means the price later moves toward the model; equal slopes in the two mean it moves as far as the outcomes justify.

\paragraph{Forecasting the forecast} We also ask how much of the price an encoder's representation can reproduce without seeing it: a ridge regression of $\operatorname{logit} p_h$ on the head's last hidden layer (256 numbers per contract), with the penalty chosen on the last quarter of Kalshi train, fitted on Kalshi train and scored by $R^2$ on Kalshi test. The nine word-history features are the reference.

\subsection{Evaluation}
Intervals are percentile 95\% intervals from \V{nBoot} bootstrap draws that resample whole calls (each Kalshi event is one call). A difference between two forecasts is computed on the same contracts within each draw. To compare encoders we also resample the ten seeds of each, so that an interval reflects both which calls are in the test and how the heads happened to train.

\paragraph{The look-ahead premium} In the pre-2012 protocol all heads are trained on the same calls before 2012, and we measure each encoder's lead over NoLBERT-Calls-2012 on calls it may have read and on calls it cannot have read. With $B_m(S)$ the Brier score of encoder $m$'s head on test set $S$ and $m_0$ NoLBERT-Calls-2012, the premium is
\begin{equation}
\label{eq:premium}
\Pi_m = \bigl[B_{m_0}(2013\text{--}23) - B_m(2013\text{--}23)\bigr] - \bigl[B_{m_0}(2025\text{--}26) - B_m(2025\text{--}26)\bigr].
\end{equation}
An encoder that profits from having read 2013--23 has a positive premium. The reference's own quality cancels out, and the two periods are bootstrapped independently.

\paragraph{Trading} For each test contract we step through the horizons 168, 96, 72, 48 and 24 hours before the call. At each, with $q$ the combined forecast of model and price, the expected profit of buying Yes at the ask $a$ is $q - a - \phi(a)$ and that of buying No at one minus the bid $b$ is $(1-q) - (1-b) - \phi(1-b)$, where $\phi(x) = 0.07\,x(1-x)$ is Kalshi's fee per contract for a trade that takes an existing offer. The first time the larger of the two exceeds a margin, we buy one contract on that side and hold it to settlement. The margin is chosen from 2, 5, 8 and 12 cents by the $t$-statistic of the profit on the Kalshi train calls (among margins with at least 100 trades), using out-of-fold forecasts from five blocks of consecutive calls, and is then applied unchanged to the test calls. The same rule driven by the recalibrated price, or by a market-only combiner that also uses the quotes, the spread and the price's last move, tells us how much can be earned from the market's own information. Standard errors are clustered by call.

\section{Results}
\label{sec:results}

\subsection{Text, word history and the four encoders}
\label{sec:results:main}

Table~\ref{tab:main} reports the main comparison. Word history alone is a strong forecaster: a logistic regression on the nine features has a Brier score of \V{briLexSyn} on the synthetic test (AUC \V{aucLexSyn}), against \V{briRuleSyn} for the rule ``said on the last call'' and \V{briBaseSyn} for the base rate. Text alone falls short of history (\V{briTextNolSyn} with NoLBERT-Calls-2023), since two earlier transcripts say more about which topics a firm discusses than about how often its managers use a given word. Combined, they beat history. NoLBERT-Mentions brings the Brier score down to \V{briNolSyn} (AUC \V{aucNolSyn}), a difference of \V{dNolLexSyn} \V{dNolLexSynCI}, or \V{pctNolLexSyn}. The result carries over to Kalshi's contracts, whose words Kalshi chose for a particular call: \V{briNolK} for NoLBERT-Mentions against \V{briLexK} for history, a difference of \V{dNolLexK} \V{dNolLexKCI}. Every encoder improves on history on both tests, with intervals that exclude zero.

\begin{table}[tbp]
\centering
\footnotesize
\setlength{\tabcolsep}{4.2pt}
\caption{Forecasting mentions on calls that no encoder has read. Brier score (lower is better) with a 95\% interval from resampling calls, and AUC. Each head reads the firm's two previous calls through a frozen encoder adapted to calls up to 2023; the full model adds the nine word-history features. NoLBERT-Mentions is the full model with NoLBERT-Calls-2023.}
\label{tab:main}
\begin{adjustbox}{max width=\linewidth}
\begin{tabular}{@{}lrrrrrr@{}}
\toprule
& \multicolumn{3}{c}{Synthetic test, 2025--26} & \multicolumn{3}{c}{Kalshi test, Apr--Nov 2026} \\
\cmidrule(lr){2-4}\cmidrule(l){5-7}
Model & Brier & 95\% interval & AUC & Brier & 95\% interval & AUC \\
\midrule
\multicolumn{7}{l}{\textit{No text}} \\
\quad Base rate & 0.2324 & \footnotesize{[0.2299, 0.2351]} & 0.500 & 0.3232 & \footnotesize{[0.3135, 0.3329]} & 0.500 \\
\quad Said on the last call & 0.1529 & \footnotesize{[0.1507, 0.1553]} & 0.783 & 0.2289 & \footnotesize{[0.2203, 0.2381]} & 0.694 \\
\quad Word history (9 features) & 0.1255 & \footnotesize{[0.1234, 0.1274]} & 0.875 & 0.1938 & \footnotesize{[0.1863, 0.2008]} & 0.783 \\
\addlinespace\multicolumn{7}{l}{\textit{Text only: the head reads the two previous calls, 3 seeds}} \\
\quad NoLBERT-Calls-2023 & 0.1702 & \footnotesize{[0.1678, 0.1724]} & 0.795 & 0.2873 & \footnotesize{[0.2759, 0.2984]} & 0.641 \\
\quad FinBERT-Calls-2023 & 0.1703 & \footnotesize{[0.1679, 0.1725]} & 0.796 & 0.2851 & \footnotesize{[0.2750, 0.2954]} & 0.628 \\
\quad DeBERTa-Calls-2023 & 0.1735 & \footnotesize{[0.1711, 0.1758]} & 0.790 & 0.2905 & \footnotesize{[0.2803, 0.3012]} & 0.630 \\
\quad ChronoBERT-Calls-2023 & 0.1730 & \footnotesize{[0.1708, 0.1751]} & 0.783 & 0.2790 & \footnotesize{[0.2687, 0.2891]} & 0.636 \\
\addlinespace\multicolumn{7}{l}{\textit{Text and word history (NoLBERT-Mentions and its rivals), 10 seeds}} \\
\quad NoLBERT-Calls-2023 & \topC{0.1221} & \footnotesize{[0.1202, 0.1240]} & \topB{0.887} & \topB{0.1893} & \footnotesize{[0.1821, 0.1960]} & \topB{0.789} \\
\quad FinBERT-Calls-2023 & \topA{0.1215} & \footnotesize{[0.1196, 0.1233]} & \topA{0.889} & \topA{0.1888} & \footnotesize{[0.1819, 0.1950]} & \topA{0.790} \\
\quad DeBERTa-Calls-2023 & \topB{0.1220} & \footnotesize{[0.1202, 0.1238]} & \topC{0.887} & \topC{0.1899} & \footnotesize{[0.1829, 0.1962]} & 0.787 \\
\quad ChronoBERT-Calls-2023 & 0.1226 & \footnotesize{[0.1207, 0.1244]} & 0.885 & 0.1904 & \footnotesize{[0.1829, 0.1970]} & \topC{0.788} \\
\addlinespace\multicolumn{7}{l}{\textit{The market}} \\
\quad Kalshi price, 1 h before & --- &  & --- & 0.1626 & \footnotesize{[0.1562, 0.1691]} & 0.842 \\
\bottomrule
\end{tabular}
 
\end{adjustbox}
\tablenotes{\shadekey{} among the four full models. The synthetic test has \V{synthTest} contracts on \V{synthTestCalls} calls from 2025--26, labelled from the transcripts; the Kalshi test has \V{kTest} contracts on \V{kTestCalls} calls from April to November 2026, labelled by Kalshi's settlement. Every Kalshi column is scored on the \V{kTestPriced1} test contracts (\V{cov1}) that have a price an hour before the call; the price row is the raw mid-quote at that time.}
\end{table}

NoLBERT-Calls holds its own against encoders pretrained on modern text (Table~\ref{tab:encoders}). With ten seeds each, the four Brier scores lie within \V{briEncRangeSyn} on the synthetic test and \V{briEncRangeK} on the Kalshi test. FinBERT-Calls is nominally ahead on the synthetic test ($\Delta = \V{eFinSyn}$), but the interval that accounts for both calls and seeds, \V{eFinSynCI}, contains zero, and no difference on Kalshi's contracts is significant. Seed noise is large in this comparison. Resampling calls alone would declare FinBERT-Calls ahead (\V{eFinSynCIcalls}) and ChronoBERT-Calls behind (\V{eChrSynCIcalls}), and single seeds vary around their mean with a standard deviation of \V{seedSdSynLo}--\V{seedSdSynHi}, as much as the gaps between encoders (Appendix Figure~\ref{fig:seeds}); a ranking of these encoders from one seed would be noise. Once the price is an input (Panel C), FinBERT-, DeBERTa- and ChronoBERT-Calls come within \V{eFinPlusMaxAbs}, \V{eDebPlusMaxAbs} and \V{eChrPlusMaxAbs} of NoLBERT-Calls at every horizon.

\begin{table}[tbp]
\centering
\footnotesize
\setlength{\tabcolsep}{3.6pt}
\caption{The four encoders, full models with ten seeds each. $\Delta$ is the row's Brier score minus that of NoLBERT-Calls-2023 (negative favours the row). Its first interval resamples calls and, for each encoder, its ten seeds; the second resamples calls alone. ``Seed pairs'' is the share of the hundred pairs of single seeds in which the row has the lower Brier score.}
\label{tab:encoders}
\begin{adjustbox}{max width=\linewidth}
\begin{tabular}{@{}lrrrrrr@{}}
\toprule
& & Seed & $\Delta$ vs & \multicolumn{2}{c}{95\% interval of $\Delta$, resampling} & Seed \\
\cmidrule(lr){5-6}
Encoder & Brier & SD & NoLBERT & calls and seeds & calls only & pairs (\%) \\
\midrule
\multicolumn{7}{l}{\textit{Panel A. Synthetic test, 2025--26 calls}} \\
\quad NoLBERT-Calls-2023 (reference) & 0.1221 & 0.0007 & --- &  &  & --- \\
\quad FinBERT-Calls-2023 & 0.1215 & 0.0008 & \textminus{}0.0006 & \footnotesize{[\textminus{}0.0013, 0.0001]} & \footnotesize{[\textminus{}0.0009, \textminus{}0.0004]} & 75 \\
\quad DeBERTa-Calls-2023 & 0.1220 & 0.0007 & \textminus{}0.0001 & \footnotesize{[\textminus{}0.0008, 0.0005]} & \footnotesize{[\textminus{}0.0003, 0.0001]} & 52 \\
\quad ChronoBERT-Calls-2023 & 0.1226 & 0.0008 & +0.0005 & \footnotesize{[\textminus{}0.0002, 0.0011]} & \footnotesize{[0.0003, 0.0007]} & 42 \\
\addlinespace\multicolumn{7}{l}{\textit{Panel B. Kalshi test, contracts priced an hour before the call}} \\
\quad NoLBERT-Calls-2023 (reference) & 0.1893 & 0.0008 & --- &  &  & --- \\
\quad FinBERT-Calls-2023 & 0.1888 & 0.0012 & \textminus{}0.0006 & \footnotesize{[\textminus{}0.0022, 0.0011]} & \footnotesize{[\textminus{}0.0019, 0.0007]} & 66 \\
\quad DeBERTa-Calls-2023 & 0.1899 & 0.0011 & +0.0005 & \footnotesize{[\textminus{}0.0008, 0.0018]} & \footnotesize{[\textminus{}0.0004, 0.0015]} & 36 \\
\quad ChronoBERT-Calls-2023 & 0.1904 & 0.0009 & +0.0010 & \footnotesize{[\textminus{}0.0003, 0.0024]} & \footnotesize{[\textminus{}0.0002, 0.0022]} & 28 \\
\addlinespace\multicolumn{7}{l}{\textit{Panel C. Kalshi test, each encoder combined with the price a week before the call}} \\
\quad NoLBERT-Calls-2023 (reference) & 0.1715 & 0.0004 & --- &  &  & --- \\
\quad FinBERT-Calls-2023 & 0.1714 & 0.0003 & \textminus{}0.0002 & \footnotesize{[\textminus{}0.0007, 0.0004]} & \footnotesize{[\textminus{}0.0006, 0.0003]} & 64 \\
\quad DeBERTa-Calls-2023 & 0.1719 & 0.0002 & +0.0004 & \footnotesize{[\textminus{}0.0001, 0.0009]} & \footnotesize{[0.0000, 0.0008]} & 18 \\
\quad ChronoBERT-Calls-2023 & 0.1718 & 0.0002 & +0.0003 & \footnotesize{[\textminus{}0.0002, 0.0008]} & \footnotesize{[\textminus{}0.0002, 0.0007]} & 28 \\
\bottomrule
\end{tabular}
 
\end{adjustbox}
\tablenotes{The Brier score is that of the ten-seed average forecast; seed SD is the standard deviation of the ten single-seed scores. In Panel C each encoder's forecast is combined with the Kalshi price 168 hours before the call (Section~\ref{sec:results:market}); the reference row there is NoLBERT-Mentions + price.}
\end{table}

\subsection{Against the market, and with it}
\label{sec:results:market}

On the day of the call, the market is the better forecaster. An hour before the call the Kalshi price has a Brier score of \V{briPriceK} on the test contracts (AUC \V{aucPriceK}), against \V{briNolK} for NoLBERT-Mentions (Table~\ref{tab:main}), and the price sharpens as the call approaches, from \V{rawPrice168} on the contracts priced a week before to \V{rawPrice1} an hour before (Figure~\ref{fig:data}c). Traders can see what two earlier transcripts cannot show, such as news since the last call and the questions analysts are likely to ask.

\begin{figure}[tbp]
\centering
\includegraphics[width=\linewidth]{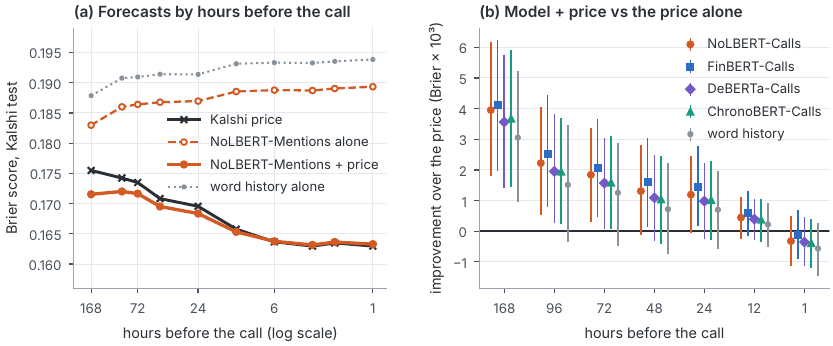}
\caption{The text and the price by hours before the call, Kalshi test. (a) Brier score of the recalibrated price, of NoLBERT-Mentions alone, of the two combined, and of word history alone, each on the test contracts priced at that horizon. (b) Improvement of each model combined with the price over the price alone, with 95\% intervals from resampling calls; positive values favour the combination.}
\label{fig:horizon}
\end{figure}

A week before the call, however, the price misses information that is in the text (Figure~\ref{fig:horizon}). Combined with NoLBERT-Mentions, the price's Brier score falls from \V{price168} to \V{nolPlus168}, a gain of \V{gainNol168} \V{gainNol168CI}, or \V{gainPctNol168}. The gain shrinks as the call approaches, to \V{gainNol96} at 96 hours and \V{gainNol72} at 72 hours, and an hour before the call it is \V{gainNol1} \V{gainNol1CI}. All four encoders share this profile: FinBERT-, DeBERTa- and ChronoBERT-Calls add \V{gainFin168}, \V{gainDeb168} and \V{gainChr168} a week before the call, with intervals that exclude zero through 72 hours for each of them. Word history adds \V{gainLex168} a week before and nothing significant from 96 hours on; the text adds more, and for longer.

\begin{figure}[tbp]
\centering
\includegraphics[width=\linewidth]{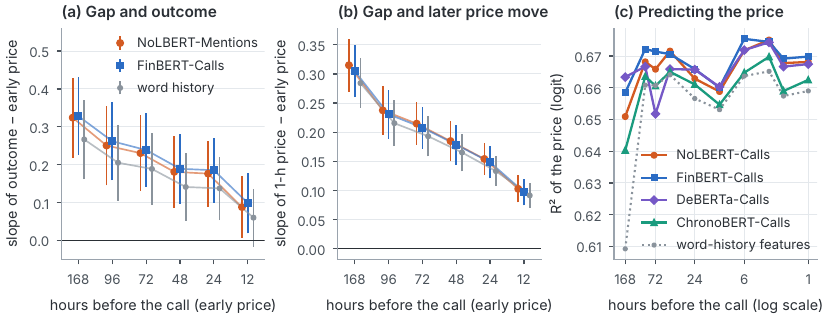}
\caption{Where the price goes, Kalshi test. (a) Slope of the outcome's surprise, $y - p_e$, and (b) of the price's later move, $p_1 - p_e$, on the model's disagreement with the early price, $q - p_e$, by early horizon $e$, with 95\% intervals clustered by call. (c) Out-of-sample $R^2$ of a ridge regression of the logit price on each encoder's head representation, or on the nine word-history features.}
\label{fig:discovery}
\end{figure}

The gain fades because the market learns what the text knew (Figure~\ref{fig:discovery}). Where NoLBERT-Mentions disagrees with the price a week before the call, the price has moved \V{slopeMoveNol168} (SE \V{seMoveNol168}) of the gap toward the model by the hour of the call (panel b). The outcomes justify a move of the same size: regressing the outcome's surprise on the same gap gives \V{slopeOutNol168} (SE \V{seOutNol168}; panel a). The market ends up where the model pointed, by as much as the model was right. Both slopes fall as the early horizon approaches the call, to \V{slopeMoveNol12} and \V{slopeOutNol12} twelve hours out. FinBERT-Calls gives the same picture; word history gives slightly flatter slopes (\V{slopeMoveLex168} and \V{slopeOutLex168} a week out).

Panel (c) turns the question around and asks how much of the price the model can reproduce without seeing it. A ridge regression on the head's hidden layer explains, out of sample, \V{rsqNol168} of the variance of the logit price a week before the call with NoLBERT-Calls, against \V{rsqLex168} for the nine word-history features; the encoders differ little. Closer to the call the history features catch up (\V{rsqLex1} against \V{rsqNol1} an hour before). The week-ahead price is largely the market's own reading of the same history and text, and what it misses, it learns before the call.

\subsection{Which words}
\label{sec:results:words}

\begin{figure}[tbp]
\centering
\includegraphics[width=\linewidth]{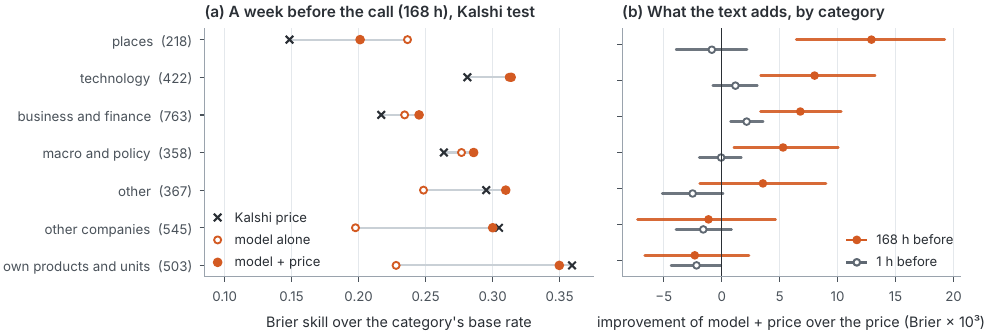}
\caption{Which words the text helps with, Kalshi test. (a) Brier skill over each category's base rate (higher is better) of the price, of NoLBERT-Mentions alone and of the two combined, a week before the call; the number of contracts is in parentheses. (b) Improvement of NoLBERT-Mentions + price over the price alone, a week and an hour before the call, with 95\% intervals from resampling calls. People (\V{nCatPerson} words) have too few contracts to show.}
\label{fig:words}
\end{figure}

The text's contribution differs across words. We assign each of the \V{nWords} words to one of eight categories from its Kalshi title and the companies that list it (Appendix Table~\ref{tab:categories}) and repeat the comparison within each (Figure~\ref{fig:words}). A week before the call, NoLBERT-Mentions alone is a better forecaster than the price for places, that is countries, regions and cities (Brier \V{catPlace168Model} against \V{catPlace168Price}), and adding it to the price helps most there, by \V{catPlace168Gain} \V{catPlace168GainCI}; for places the model alone even beats the combination, whose weights are fitted on all words. Technology (\V{catTech168Gain}), business and finance (\V{catBus168Gain}) and macro and policy (\V{catMacro168Gain}) follow, all with intervals that exclude zero. For the firm's own products and units and for other companies, the price is far better than the model (\V{catOwn168Price} against \V{catOwn168Model} for own products) and the text adds nothing (\V{catOwn168Gain} and \V{catOtherCo168Gain}). A natural reading is that whether executives mention a country or a technology depends on what the firm has been discussing, which is in its previous calls, while whether they mention a new product or a rival depends on events since the last call, which traders follow and the model cannot see. An hour before the call only business and finance keeps a gain (\V{catBus1Gain} \V{catBus1GainCI}).

\subsection{What a backtest would have shown}
\label{sec:results:hindsight}

\begin{figure}[tbp]
\centering
\includegraphics[width=\linewidth]{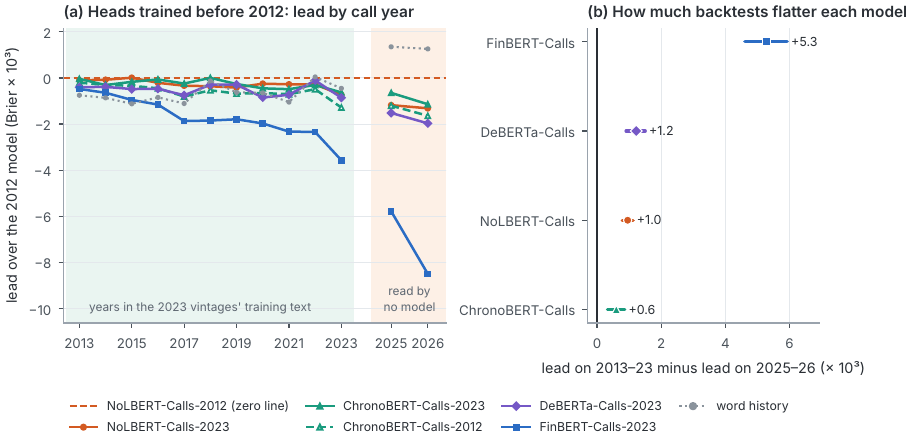}
\caption{How much backtests flatter each encoder. Heads are trained on calls before 2012. (a) Lead of each head over NoLBERT-Calls-2012 by call year (Brier score of NoLBERT-Calls-2012 minus the head's; positive favours the head). The shaded areas are the years the 2023 encoders read during adaptation and the years no encoder has read. (b) The look-ahead premium of Equation~\eqref{eq:premium}, with 95\% intervals.}
\label{fig:lookahead}
\end{figure}

Kalshi's test is free of hindsight for all four encoders, since every test call postdates their text. A backtest on the long record of calls is not: the 2023 encoders read the calls of 2013--23 during adaptation, and FinBERT's pretraining read calls from 2004--19. To measure how much this flatters each encoder, we train heads on calls before 2012, tune them on 2012 and score them on calls from 2013--23 and from 2025--26 (Table~\ref{tab:lookahead} and Figure~\ref{fig:lookahead}).

\begin{table}[tbp]
\centering
\footnotesize
\setlength{\tabcolsep}{5pt}
\caption{Heads trained before 2012. Brier score on synthetic contracts from calls of 2013--23, which the 2023 encoders read during adaptation, and of 2025--26, which no encoder read, and on the Kalshi test. The premium (Equation~\ref{eq:premium}) is the encoder's lead over NoLBERT-Calls-2012 on 2013--23 minus its lead on 2025--26, with a 95\% interval; a positive premium means a backtest flatters the encoder.}
\label{tab:lookahead}
\begin{adjustbox}{max width=\linewidth}
\begin{tabular}{@{}llrrrr@{}}
\toprule
& Text up to & 2013--23 & 2025--26 & Kalshi test & Premium vs NoLBERT-Calls-2012 \\
\midrule
NoLBERT-Calls-2012 (reference) & 2012 & \topA{0.0876} & \topB{0.1279} & \topA{0.1971} & --- \\
ChronoBERT-Calls-2012 & 2012 & 0.0882 & 0.1293 & 0.2001 & --- \\
NoLBERT-Calls-2023 & 2023 & \topC{0.0879} & 0.1292 & \topC{0.1992} & +0.0010 \footnotesize{[0.0008, 0.0011]} \\
ChronoBERT-Calls-2023 & 2023 & \topB{0.0879} & \topC{0.1288} & \topB{0.1984} & +0.0006 \footnotesize{[0.0003, 0.0009]} \\
DeBERTa-Calls-2023 & 2023$^{\dagger}$ & 0.0881 & 0.1296 & 0.2000 & +0.0012 \footnotesize{[0.0009, 0.0015]} \\
FinBERT-Calls-2023 & 2023$^{\ddagger}$ & 0.0894 & 0.1350 & 0.2137 & +0.0053 \footnotesize{[0.0046, 0.0059]} \\
Word history (9 features) & 2011 & 0.0883 & \topA{0.1266} & --- & --- \\
\bottomrule
\end{tabular}
 
\end{adjustbox}
\tablenotes{\shadekey. Full models, three seeds each. $^{\dagger}$DeBERTa's pretraining text has no date control. $^{\ddagger}$FinBERT's pretraining corpus includes earnings calls from 2004--19. Adaptation: NoLBERT-Calls-2012, 480 minutes on calls from all countries; ChronoBERT-Calls-2012, 150 minutes on US calls. Word history is trained on the same calls before 2012.}
\end{table}

NoLBERT-Calls-2012, which has read nothing written after 2012, is the best or tied encoder in every test year, including the years the 2023 encoders read; in no year does another head's Brier score fall below it by as much as 0.0001. It also leads ChronoBERT-Calls-2012, the other encoder that stops in 2012, on every test set: by \V{chrTwVsNolTwSeen} \V{chrTwVsNolTwSeenCI} on 2013--23, \V{chrTwVsNolTwUnseen} on 2025--26 and \V{chrTwVsNolTwK} \V{chrTwVsNolTwKCI} on the Kalshi test. The look-ahead premium is small for NoLBERT-Calls-2023 (\V{premNol} \V{premNolCI}), ChronoBERT-Calls-2023 (\V{premChr}) and DeBERTa-Calls-2023 (\V{premDeb}). For FinBERT-Calls-2023 it is \V{premFin} \V{premFinCI}, \V{premRatioFin} times NoLBERT-Calls'. Figure~\ref{fig:lookahead}a shows the pattern behind it: the FinBERT-Calls head falls further behind the 2012 reference almost every year, from \V{finGap2013} in 2013 to \V{finGap2023} in 2023, and then to \V{finGap2025} and \V{finGap2026} in the years no encoder has read. A backtest on 2013--23 would place FinBERT-Calls far closer to the other encoders than it is on new calls.

The pre-2012 heads also show why a mention model has to be retrained. By 2025--26, word history trained on the same calls from 2006--11 beats every one of them (by \V{lexBeatsNolTwUnseen} against NoLBERT-Calls-2012), although the text heads lead on 2013--23; the main protocol's heads, trained through 2023, keep a clear lead over history (Section~\ref{sec:results:main}).

\subsection{A trading rule}
\label{sec:results:trading}

\begin{figure}[tbp]
\centering
\includegraphics[width=\linewidth]{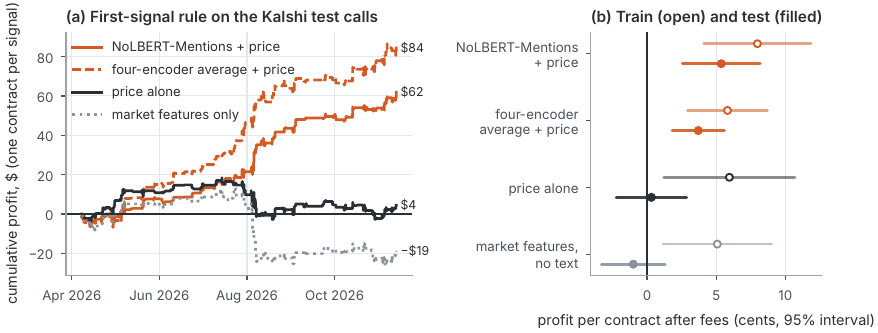}
\caption{The first-signal rule. (a) Cumulative profit on the Kalshi test calls, one contract per signal, after fees. (b) Profit per contract after fees on Kalshi train (out-of-fold forecasts, open markers) and on Kalshi test (filled markers), with 95\% intervals clustered by call.}
\label{fig:trading}
\end{figure}

The week-ahead information is worth money after Kalshi's fees. The first-signal rule of Section~\ref{sec:methods} buys one contract on the side the combined forecast favours the first time, between a week and a day before the call, that its expected profit after the fee exceeds a margin chosen on the training calls. Table~\ref{tab:trading} and Figure~\ref{fig:trading} report the result on the test calls.

\begin{table}[tbp]
\centering
\footnotesize
\setlength{\tabcolsep}{3.6pt}
\caption{The first-signal trading rule, after Kalshi's fees. The rule buys one contract on the favoured side the first time, from 168 to 24 hours before the call, that the expected profit exceeds the margin, and holds it to settlement. The margin is chosen on Kalshi train by the $t$-statistic of out-of-fold profits.}
\label{tab:trading}
\begin{adjustbox}{max width=\linewidth}
\begin{tabular}{@{}lrrrrrrrrr@{}}
\toprule
& Margin & Train, out of fold & \multicolumn{7}{c}{Kalshi test} \\
\cmidrule(lr){3-3}\cmidrule(l){4-10}
Probability used & (cents) & cents ($t$) & cents (SE) & $t$ & Contracts & Calls & Profit (\$) & Return (\%) & Hit (\%) \\
\midrule
NoLBERT-Mentions + price & 5 & +8.0 (4.0) & +5.4 (1.4) & 3.7 & 1,154 & 291 & 62 & 9 & 63 \\
Four-encoder average + price & 2 & +5.8 (3.9) & +3.7 (1.0) & 3.8 & 2,277 & 316 & 84 & 6 & 64 \\
Price alone (recalibrated) & 5 & +6.0 (2.5) & +0.3 (1.3) & 0.2 & 1,425 & 292 & 4 & 0 & 65 \\
Market features, no text & 5 & +5.1 (2.5) & \textminus{}1.0 (1.2) & \textminus{}0.8 & 1,876 & 294 & \textminus{}19 & \textminus{}2 & 58 \\
\bottomrule
\end{tabular}
 
\end{adjustbox}
\tablenotes{Profit per contract in cents with standard errors clustered by call. Return is total profit over the total cost of the contracts bought, fees included; hit is the share of profitable trades. The price-alone and market rows are placebos: the market-only forecast also uses the quotes, the spread and the price's last move, but no text.}
\end{table}

With NoLBERT-Mentions + price, the rule trades \V{trNolN} contracts on \V{trNolEvents} of the \V{kTestCalls} test calls and earns \V{trNolCents} cents per contract after fees (SE \V{trNolSe}, $t = \V{trNolT}$): \V{trNolDollars} on \V{trNolCapital} spent, a return of \V{trNolRoc}, with \V{trNolHit} of trades profitable. The four-encoder average, at a lower margin, trades twice as often and earns \V{trAvgCents} cents ($t = \V{trAvgT}$). Without text the rule earns nothing: \V{trPriceCents} cents on the recalibrated price ($t = \V{trPriceT}$) and \V{trMktCents} on the market-only forecast ($t = \V{trMktT}$), although both looked profitable on the training calls (\V{trPriceTrainCents} and \V{trMktTrainCents} cents out of fold).

The profit is broad (Appendix~\ref{app:trading}). It is positive in every quarter of the test period (\V{trNolQtwoCents}, \V{trNolQthreeCents} and \V{trNolQfourCents} cents) and on both sides (\V{trNolNoCents} cents on the \V{trNolNoN} purchases of No, \V{trNolYesCents} on the \V{trNolYesN} purchases of Yes), and it is unchanged when fees are rounded up to the cent, as Kalshi charges them (\V{trNolRoundCents}). We chose the rule's settings on the training calls within each of three families of rules, and the first-signal family also had the highest training $t$-statistic (\V{trNolFirstTrainT}, against \V{trNolSettleTrainT} and \V{trNolExitTrainT}); on the test calls, holding from a single horizon earns \V{trNolSettleTestCents} cents ($t = \V{trNolSettleTestT}$) and selling at the quote an hour before the call \V{trNolExitTestCents} ($t = \V{trNolExitTestT}$). Every rule trades one contract per signal at the quoted price.

\section{Conclusion}
\label{sec:discussion}

An encoder that has read nothing written after 2023 can forecast what managers will say on their earnings calls, and a week before the call it knows something a prediction market does not. NoLBERT-Mentions improves on the word's own history by \V{pctNolLexSyn} on calls no encoder has read. Combined with the Kalshi price, it lowers the price's Brier score by \V{gainPctNol168} a week before the call, and a rule that trades on the disagreement earns \V{trNolCents} cents per contract after fees on calls the model never saw. The market closes the gap over the following days, moving toward the model by as much as the outcomes justify, so that by the hour of the call the price contains what the text knew: in the terms of \citet{page2013calibrated}, the market is accurate close to expiry and less so a week away.

For research with language models, the lesson concerns backtests. Encoders adapted to the same calls are interchangeable as forecasters of mentions, but their backtests are not: a backtest on the years it read flatters FinBERT-Calls \V{premRatioFin} times as much as NoLBERT-Calls, and NoLBERT-Calls-2012 is the best or tied encoder in every test year from 2013 to 2026. A researcher who wants to know how a text-based forecast would have done in the past gives up no measurable accuracy, and avoids hindsight, by reading the calls with NoLBERT. Together with NoLBERT-Calls \citep{foley2026nolbertcallsreadingearningscalls,prakash2026nolbertcalls}, which builds the reader, these results make the case for hindsight-free encoders in finance: the same model reads past calls accurately and anticipates the next one.

As Kalshi's markets grow beyond their first seasons, the same design will give sharper estimates by word category and horizon, and it carries over directly to Kalshi's mention markets on Federal Reserve press conferences and political speeches. The surprise in a mention, measured against the price an hour before the call, is also a new variable for studying how stock prices react to what managers say.

\clearpage
\bibliographystyle{plainnat}
\bibliography{references}

\clearpage
\appendix
\setcounter{table}{0}
\renewcommand{\thetable}{A\arabic{table}}
\setcounter{figure}{0}
\renewcommand{\thefigure}{A\arabic{figure}}

\section{Implementation details}
\label{app:details}

Table~\ref{tab:hyper} lists the settings of the head and its training. The head's settings were fixed before any test set was scored; the only choice made on data is the learning rate, chosen from two values by the validation loss of seed 0 and then used for every seed.

\begin{table}[htbp]
\centering
\footnotesize
\caption{Settings of NoLBERT-Mentions and of the rival heads.}
\label{tab:hyper}
\begin{tabular}{@{}lp{0.74\linewidth}@{}}
\toprule
Setting & Value \\
\midrule
Context & the firm's two previous calls, up to 64 passages each (256 tokens), typed by call and section \\
Word query & ``What can you tell us about \emph{word}?'' (128 tokens) \\
Encoders & frozen; mean-pooled, unit-length embeddings \\
Head & projection to 256, scaled dot-product attention from the word to the passages, 7 similarity statistics, MLP 256--256 (GELU, dropout 0.1) \\
Output & negative binomial count of mentions; P(said at least the contract's threshold) \\
Loss & negative binomial log likelihood (count capped at 50) + binary cross-entropy \\
Word-history features & said last call, log count last call, share said in the last 4 and 8 calls, said in any of the last 8, analysts said it last call, share of all calls in the past 30 days that said it, count threshold, length of history \\
Optimisation & AdamW, weight decay 0.01, 32 calls per batch, up to 8 epochs, early stopping after 2 \\
Learning rate & chosen from \{0.001, 0.0003\} by validation loss with seed 0, then used for every seed \\
Seeds & 3 per run (protocol); 10 for the main runs of the four 2023 encoders \\
Intervals & 1,000 cluster-bootstrap draws (calls; Kalshi events), percentile 95\% \\
Kalshi fee & 0.07 $\times$ price $\times$ (1 $-$ price) per contract (taker) \\
\bottomrule
\end{tabular}
 
\end{table}

Table~\ref{tab:categories} defines the word categories of Section~\ref{sec:results:words}. Each word is assigned from its Kalshi title and the companies that list it, by keyword rules, by a default that treats a word listed for one company as one of its own products or units unless a rule applies, and by reading the remaining titles; the assignment is part of the replication files.

\begin{table}[htbp]
\centering
\footnotesize
\caption{Word categories. ``Words'' counts the distinct words Kalshi has listed; the last column counts the Kalshi test contracts priced an hour before the call.}
\label{tab:categories}
\begin{tabular}{@{}lp{0.52\linewidth}rr@{}}
\toprule
Category & What it holds & Words & Kalshi test contracts \\
\midrule
Own products and units & the firm's products, brands, segments, programmes, acquisitions & 796 & 713 \\
Business and finance & margins, buybacks, guidance, pricing, demand, costs & 513 & 1,096 \\
Other companies & named firms other than the firm itself & 394 & 746 \\
Technology & AI, cloud, GPUs, electric vehicles, crypto (not one firm's brand) & 326 & 578 \\
Macro and policy & tariffs, inflation, rates, regulation, government, shutdowns & 139 & 538 \\
Places & countries, regions, cities & 117 & 330 \\
People & named people & 32 & 43 \\
Other & everything else & 324 & 537 \\
\bottomrule
\end{tabular}
 
\end{table}

\section{Seeds}
\label{app:seeds}

\begin{figure}[htbp]
\centering
\includegraphics[width=\linewidth]{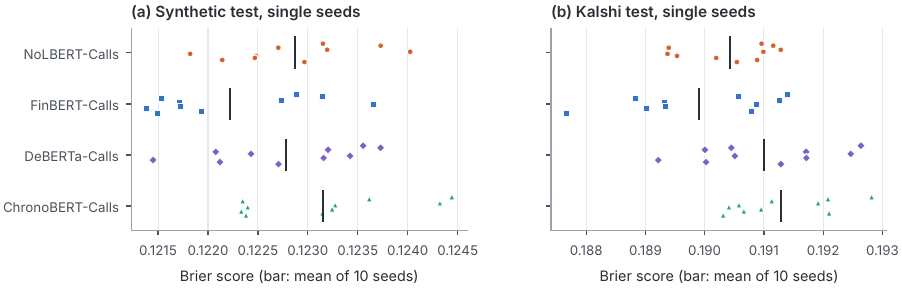}
\caption{Brier scores of single seeds, full models of the four 2023 encoders. The bar marks the mean of the ten single-seed scores; the ten-seed average forecast scores lower (Table~\ref{tab:encoders}).}
\label{fig:seeds}
\end{figure}

The four 2023 encoders' full models are trained with ten seeds each. Figure~\ref{fig:seeds} shows the Brier score of every single seed. The seeds of one encoder spread over a range as wide as the gaps between encoders: on the synthetic test, NoLBERT-Calls' single seeds range from \V{nolSeedMin} to \V{nolSeedMax} and FinBERT-Calls' from \V{finSeedMin} to \V{finSeedMax}. The ten-seed average forecast, which Tables~\ref{tab:main} and~\ref{tab:encoders} report, scores better than the typical single seed and about as well as the best one. The intervals of Table~\ref{tab:encoders} resample calls and seeds together: each draw resamples the calls and, for each encoder, ten seeds with replacement, and recomputes both ten-seed averages.

\section{Richer ways to combine text and price}
\label{app:combiners}

The two-input logistic regression of Section~\ref{sec:methods} is deliberately simple. Table~\ref{tab:combiners} compares it with richer combiners that add the mid-quote and the spread at each horizon, whether the contract is quoted, the price's move since the previous horizon, the word-history features and the word-history model's forecast, the four encoders' forecasts separately or as an average, and boosted trees with monotone constraints. Each is chosen by blocked five-fold cross-validation on the Kalshi train contracts, with folds of consecutive calls, and the out-of-fold Brier score is averaged over the horizons. The selected combiner is the average of the four encoders' forecasts plus the price (\V{cvSelectedScore}), tied with NoLBERT-Mentions plus the price (\V{cvNolScore}); the price alone scores \V{cvPriceScore}. Every richer combiner does worse in cross-validation, and on the test calls the market-only combiners add nothing to the price. The selected combiner gains \V{cmbAvg168} over the price a week before the call, the same as NoLBERT-Mentions + price (\V{cmbNol168}). Because each combiner here is fitted to the ten-seed average forecast, with its regularisation chosen by cross-validation, its gains differ in the fourth decimal from those of Figure~\ref{fig:horizon}, which combine each seed with the price and average the results.

\begin{table}[htbp]
\centering
\footnotesize
\setlength{\tabcolsep}{4pt}
\caption{Combiners of text and market information, ordered by cross-validated Brier score on Kalshi train. Columns 3 to 8 give the improvement over the recalibrated price on the Kalshi test (the price row gives its Brier score).}
\label{tab:combiners}
\begin{adjustbox}{max width=\linewidth}
\begin{tabular}{@{}lrrrrrrr@{}}
\toprule
& Train CV & \multicolumn{6}{c}{Kalshi test: gain over the price, by hours before the call} \\
\cmidrule(lr){2-2}\cmidrule(l){3-8}
Combiner & Brier & 168 & 96 & 72 & 48 & 24 & 1 \\
\midrule
four-encoder average + price $\star$ & 0.14705 & +0.0042$^{*}$ & +0.0024$^{*}$ & +0.0018$^{*}$ & +0.0012 & +0.0011 & \textminus{}0.0004 \\
NoLBERT-Mentions + price & 0.14706 & +0.0043$^{*}$ & +0.0024$^{*}$ & +0.0018$^{*}$ & +0.0012 & +0.0011 & \textminus{}0.0004 \\
FinBERT-Calls + price & 0.14710 & +0.0044$^{*}$ & +0.0027$^{*}$ & +0.0020$^{*}$ & +0.0015$^{*}$ & +0.0014$^{*}$ & \textminus{}0.0002 \\
four encoders + price & 0.14728 & +0.0037$^{*}$ & +0.0022$^{*}$ & +0.0020$^{*}$ & +0.0015 & +0.0014$^{*}$ & \textminus{}0.0002 \\
average + market & 0.14739 & +0.0029$^{*}$ & \textminus{}0.0013 & +0.0018 & +0.0013 & +0.0005 & \textminus{}0.0002 \\
price & 0.14819 & 0.1754 & 0.1741 & 0.1734 & 0.1708 & 0.1695 & 0.1629 \\
market & 0.14849 & \textminus{}0.0008 & \textminus{}0.0027$^{*}$ & +0.0006 & +0.0000 & \textminus{}0.0004 & \textminus{}0.0001 \\
average + market + history & 0.14853 & +0.0017 & \textminus{}0.0009 & +0.0018 & +0.0006 & +0.0004 & +0.0002 \\
boosted trees: market + history & 0.14859 & \textminus{}0.0014 & \textminus{}0.0064$^{*}$ & +0.0012 & \textminus{}0.0007 & \textminus{}0.0020$^{*}$ & \textminus{}0.0022$^{*}$ \\
market + history & 0.14864 & +0.0006 & \textminus{}0.0011 & +0.0019 & +0.0003 & +0.0002 & \textminus{}0.0001 \\
boosted trees: average + market + history & 0.14881 & +0.0002 & \textminus{}0.0056$^{*}$ & +0.0015 & \textminus{}0.0006 & \textminus{}0.0015 & \textminus{}0.0022$^{*}$ \\
\bottomrule
\end{tabular}
 
\end{adjustbox}
\tablenotes{$\star$ marks the combiner selected by cross-validation; $^{*}$ marks a test improvement whose 95\% interval from resampling calls excludes zero. ``Market'' combiners use the price, the mid-quote, the spread, whether the contract is quoted and the price's last move; ``history'' adds the nine word-history features and the word-history model's forecast; ``average'' is the average of the four encoders' forecasts; boosted trees are LightGBM with monotone constraints on the probabilities.}
\end{table}

\section{Trading}
\label{app:trading}

Table~\ref{tab:trading-slices} splits the test trades of the first-signal rule with NoLBERT-Mentions + price. Table~\ref{tab:trading-families} reports all three families of rules we considered, each with its horizon and margin chosen on the Kalshi train calls by the $t$-statistic of out-of-fold profits: the first-signal rule; buying at a single horizon and holding to settlement; and buying at a single horizon and selling at the quote an hour before the call. For the price alone and the market-only forecast, the family with the best training $t$-statistic is the single-horizon rule, which earns \V{trPriceBestCents} and \V{trMktBestCents} cents on the test calls ($t = \V{trPriceBestT}$ and \V{trMktBestT}). Selling before the call loses money for both forecasts without text, because the round trip pays the spread and the fee twice; with text it roughly breaks even.

\begin{table}[htbp]
\centering
\footnotesize
\caption{The first-signal rule with NoLBERT-Mentions + price on the Kalshi test calls, by slice.}
\label{tab:trading-slices}
\begin{tabular}{@{}lrrrrr@{}}
\toprule
Slice of the Kalshi test trades & Contracts & Calls & Cents & SE & $t$ \\
\midrule
Calls in Apr--Jun 2026 & 346 & 81 & +3.5 & 2.7 & 1.3 \\
Calls in Jul--Sep 2026 & 552 & 113 & +7.0 & 2.1 & 3.4 \\
Calls in Oct--Nov 2026 & 256 & 97 & +4.4 & 3.1 & 1.4 \\
Bought NO & 1,070 & 283 & +4.5 & 1.5 & 3.0 \\
Bought YES & 84 & 61 & +16.4 & 4.6 & 3.5 \\
Entered 168 h before & 485 & 176 & +6.6 & 2.2 & 3.0 \\
Entered 96 h before & 226 & 127 & +4.9 & 3.2 & 1.5 \\
Entered 72 h before & 188 & 128 & +5.8 & 3.1 & 1.9 \\
Entered 48 h before & 143 & 102 & +1.4 & 3.3 & 0.4 \\
Entered 24 h before & 112 & 79 & +5.1 & 4.3 & 1.2 \\
Fees rounded up to the cent & 1,034 & 282 & +5.4 & 1.5 & 3.6 \\
\bottomrule
\end{tabular}
 
\tablenotes{Profit per contract in cents after fees, with standard errors clustered by call. The quarters are those of the call. The last row charges each contract's fee rounded up to the next cent.}
\end{table}

\begin{table}[htbp]
\centering
\footnotesize
\setlength{\tabcolsep}{4pt}
\caption{Three families of trading rules, with the horizon and margin of each chosen on Kalshi train.}
\label{tab:trading-families}
\begin{adjustbox}{max width=\linewidth}
\begin{tabular}{@{}lllrrrr@{}}
\toprule
& & Horizon, margin & Train, out of fold & \multicolumn{3}{c}{Kalshi test} \\
\cmidrule(lr){4-4}\cmidrule(l){5-7}
Probability used & Rule & (cents) & cents ($t$) & cents (SE) & $t$ & Contracts \\
\midrule
NoLBERT-Mentions + price & First signal, 168--24 h & 5 & +8.0 (4.0) & +5.4 (1.4) & 3.7 & 1,154 \\
 & One horizon, to settlement & 48 h, 5 & +8.6 (3.8) & +3.7 (1.7) & 2.2 & 763 \\
 & One horizon, sold 1 h before & 168 h, 8 & \textminus{}2.7 (\textminus{}1.9) & +0.9 (1.4) & 0.7 & 265 \\
\addlinespace
Four-encoder average + price & First signal, 168--24 h & 2 & +5.8 (3.9) & +3.7 (1.0) & 3.8 & 2,277 \\
 & One horizon, to settlement & 72 h, 2 & +6.7 (3.8) & +2.8 (1.1) & 2.5 & 1,449 \\
 & One horizon, sold 1 h before & 168 h, 8 & \textminus{}3.0 (\textminus{}2.2) & +1.1 (1.4) & 0.8 & 272 \\
\addlinespace
Price alone & First signal, 168--24 h & 5 & +6.0 (2.5) & +0.3 (1.3) & 0.2 & 1,425 \\
 & One horizon, to settlement & 24 h, 5 & +11.2 (3.1) & \textminus{}0.3 (1.8) & \textminus{}0.2 & 754 \\
 & One horizon, sold 1 h before & 168 h, 5 & \textminus{}7.1 (\textminus{}5.6) & \textminus{}5.3 (1.0) & \textminus{}5.6 & 358 \\
\addlinespace
Market features, no text & First signal, 168--24 h & 5 & +5.1 (2.5) & \textminus{}1.0 (1.2) & \textminus{}0.8 & 1,876 \\
 & One horizon, to settlement & 24 h, 2, sized by edge & +5.6 (2.9) & +1.2 (1.2) & 1.0 & 5,699 \\
 & One horizon, sold 1 h before & 72 h, 5 & \textminus{}5.8 (\textminus{}6.4) & \textminus{}6.5 (0.8) & \textminus{}7.8 & 660 \\
\bottomrule
\end{tabular}
 
\end{adjustbox}
\tablenotes{Profit per contract in cents after fees. Train profits use out-of-fold forecasts from blocked cross-validation; the $t$-statistic in parentheses is the selection criterion (at least 100 trades). ``Sized by edge'' buys one contract per two cents of expected profit, up to ten.}
\end{table}

\end{document}